\documentclass[a4paper,12pt]{article}
\usepackage{graphicx}
\newcommand{\be}{\begin{equation}}
\newcommand{\ee}{\end{equation}}
\newcommand{\bea}{\begin{eqnarray}}
\newcommand{\eea}{\end{eqnarray}}
\newcommand{\nn}{\nonumber}

\newcommand{\noi}{\noindent}

\begin{document}
\thispagestyle{empty}
\vspace*{-2cm}

\bigskip

\begin{center}

  {\large\bf
 
A random walk with heavy flavours

\vspace{1.5cm}
  }
\medskip
  {\large

 Surasree Mazumder$^{\dag}$\footnote{surasree.mazumder@gmail.com}, 
Trambak Bhattacharyya$^{\dag}$\footnote{trambak.bhattacharyya@gmail.com}
and Santosh K. Das$^{*}$\footnote{dsantoshphy@gmail.com}}
  \bigskip

{\it

$^{\dag}$ Variable Energy Cyclotron Center, 1/AF, Bidhan Nagar, Kolkata-700064, India\\

$^{*}$ Department of Physics and Astronomy, University of Catania, Via S. Sofia 64, I-
95125 Catania, Italy\\

\begin{abstract}
We focus on evaluating transport coefficients
like drag and diffusion of heavy quarks (HQ) passing  
through Quark Gluon Plasma using perturbative QCD (pQCD).
Experimental observable like nuclear suppression factor
($R_{AA}$) of HQ is evaluated for 
both zero and non-zero baryonic chemical potential ($\mu_B$) scenarios
using Fokker-Planck equation. Theoretical estimates of $R_{AA}$ 
are contrasted with experiments. 
\end{abstract}

\vspace{2.7cm}}

\end{center}

\vspace{1.0cm}

\section{Introduction:}

When  nuclear matter is subjected to an ambience of very high density, the 
individual quarks and gluons would no longer be confined within the hadrons but melt into a deconfined 
state of quarks and gluons. Just after the discovery of asymptotic freedom~\cite{Gross,Gross1,Politzer}, 
Collins and Perry~\cite{Collins} also suggested that at very high density the degrees of freedom of the strongly 
interacting matters are not hadrons but quarks and gluons. 
The same is true when QCD vacuum is excited to high temperatures, too (\cite{parisi}).
With increasing temperature, new and new hadrons are produced thereby increasing the corresponding
number density; and at a certain temperature, there is an overlap of hadrons. Such a phase of matter 
is called Quark Gluon Plasma (QGP) and its study needs QCD, the theory of strong interaction 
which is extremely successful in vacuum, to be applied in a thermal medium. So, the deconfined 
state of quarks and gluons gives an opportunity to peruse `condensed matter 
physics' of elementary particles in the new domain of non-abelian gauge theory. 

Lattice QCD based calculations predict that the typical value of the temperature for the 
quark-hadron transition, $T_c$, is $\sim$170 MeV~\cite{Karsch0}\footnote{Latest lattice QCD 
results show that $T_c\sim160$ MeV \cite{borsanyi,nature,wupparteljpg}}. According to the cosmological 
big bang model the universe has undergone several phase transitions (GUT, 
Electroweak, quark to hadron etc) at different stages of its evolution. The 
quark-hadron transition occurred  when the universe was few 
microseconds ($\mu s$) old and this is the only transition which can be accessed in the 
laboratory currently. The study of quark-hadron transition 
demands special importance in understanding the evolution of the $\mu s$ old early 
universe. The issue is very crucial for astrophysics too, as the 
core of the compact astrophysical objects like neutron stars may contain 
quark matter at high baryon density and low temperature. 
So there is a multitude of reasons behind creating QGP in laboratories.

Temperature and energy density required to 
produce QGP in the laboratory can be achieved by colliding heavy ions at relativistic energies,  
under controlled laboratory environment. The nuclear collisions at Relativistic Heavy Ion Collider (RHIC)
and the Large Hadron Collider (LHC) energies (200 GeV/A and 2.76 TeV/A respectively) 
are aimed at creating QGP. Once the QGP medium is created,
we must try to understand the transport properties of the medium {\it i.e.} whether it is liquid or gas.
The study of the transport coefficients of strongly correlated system is a field of
high contemporary interest both theoretically and experimentally. In one hand, the calculation of
the lower bound on the shear viscosity ($\eta$) to entropy density ($s$)  ratio ($\eta/s$) within the 
frame work of AdS/CFT model~\cite{kss}  has ignited enormous interests among the theorists. On the 
other hand, the experimental study of the $\eta/s$ for cold atomic systems and QGP 
and their similarities have generated huge interest across various branches of 
physics (see~\cite{aadams} for a review).

In general, the interaction of probes with a medium  brings out useful information about the
nature of the medium. As the magnitude of the transport coefficients are sensitive to 
the coupling strength, so these quantities qualify as useful quantities to characterize 
a  medium.

In context of probing QGP, the heavy quarks (HQs), mainly, 
charm and bottom quarks, play a vital role. The reasons are as follows:

\begin{table}[ht]
\centering
\begin{tabular}{llc}
\\
\\
$\bullet$ HQ mass is significantly larger than the typically attained temperatures \\
and other nonperturbative scales, $M \gg T_c, \Lambda_{QCD}$(intrinsic energy scale \\
for the strong interaction), {\it i.e.} the production of HQs is essentially constrained \\
to the early, primordial stage of a heavy-ion collision and they do not dictate the bulk \\
properties of the matter. Therefore, the heavy flavours are the witness to the entire \\
space-time evolution of the system.
\\
\\
$\bullet$ Their thermalization time scale is larger by a factor of $m/T$, where $m$ \\
is the mass of heavy quarks and $T$ is the temperature, than that of the light quarks \\
and gluons and hence heavy quarks can retain the interaction history very effectively.
\\

\end{tabular}
\label{symop}
\end{table}



From experimental point of view, however, the issue of HQ thermalization in QGP can be 
addressed by measuring the elliptic flow ($v_2$) of leptons from the decays
of HQs. Moreover, the  observed transverse momentum suppression ($R_{AA}$) of 
leptons originating from the decays of
D and B mesons produced in nuclear 
collisions as compared to those produced in proton+proton (pp) collisions at
the same colliding energy~\cite{phenixe,stare,alice} offer us an opportunity to estimate
the drag and diffusion coefficients of QGP\footnote{It is now possible to detect directly 
D mesons at LHC detectors like ALICE, see \cite{ALICE}}.
Hence, no wonder that in the recent past large 
number of attempts have been made to study both heavy flavour suppression~\cite{phenixe,stare} 
and elliptic flow~\cite{phenixelat} within the framework of perturbative 
QCD (pQCD)~\cite{moore,ko,adil,urs,gossaich,das,san,dams,gre,alberico,mbad,younus,mustafa,jeon,bass,hees,dca}.

\section{Motion of Heavy Quarks (HQs) in QGP:}
In the introduction we have already discussed that HQs act as effective probes to look into the 
properties of QGP. As HQs are much heavier than the particles constituting the QGP
thermal bath, one expects that they will execute brownian motion in QGP medium {\cite{syam,arsprl}}. 
The system under study, then, would have two components: (i) the QGP formed at an initial 
temperature $T_i$ and initial thermalization time $\tau_i$ consisting of light quarks and gluons 
and (ii) Heavy Quark, the Brownian particle formed due to hard collisions at very early stage of heavy ion collision.
The momentum distribution of HQ is governed by a non-linear integro differential equation
which is the Boltzman Transport Equation (BTE):
\bea
\left[\frac{\partial}{\partial t}+\frac{\textbf{p}}{E}.\frac{\partial}{\partial \textbf{x}}
+\textbf{F}.\frac{\partial}{{\partial \textbf{p}}}
\right] f(\textbf{x},\textbf{p},t)=\left[\frac{\partial f}{\partial t}\right]_{collisions}
\label{BTE}
\eea
\textbf{F} is the force exerted on the HQ by the surrounding colour field. \textbf{p} and E
denote the three momentum and the energy of the HQ respectively. The right hand side of 
Eq.~\ref{BTE}, which is called the collision  integral, $C[f]$, is attributed to the 
QCD interactions of HQ with light quarks ($q$), anti-quarks ($\bar{q}$) and
gluons (g). One should, in principle, solve this differential equation under the influence of 
potential involving interaction of HQs with light quarks/anti-quarks
and the background colour field in the force term.
But, here, we will set $\textbf{F}=0$ and will treat
QGP to be uniform. Therefore, the second and the third term of the left hand side of
Eq.~\ref{BTE} vanish under these approximations. Again defining,
\bea
f(\textbf{p},t)= \frac{1}{V} \int d^{3}\textbf{x} f(\textbf{x},\textbf{p},t)
\eea
which is the normalized probability distribution in the momentum space, we have
\be
\frac{\partial f(\textbf{p},t)}{\partial t}=\left[\frac{\partial f}{\partial t}\right]_{collisions}.
\label{BTE1}
\ee
Eq.~\ref{BTE1} signifies that all variation of the distribution function of HQ with time
is due to the collisions only. 

\section{Formalism:}
Our main aim is to determine the collision integral of the transport Eq.~\ref{BTE1}. Once, we 
determine certain form of $C[f]$, we can proceed towards solving the differential equation.
There are lots of approximations through which the integro-differential equation
can be solved. 
Of course, under certain conditions, Eq.~\ref{BTE1} can be reduced to the simple form:
\be
\frac{\partial f(\textbf{p},t)}{\partial t}=-\frac{f-f_0}{\tau}
\ee
which is a useful first approximation. Here, $f_{0}$ is the equilibrium distribution function and 
$\tau$ is the relaxation time that determines the rate at which the fluctuations in the system
drive it to a state of equilibrium again. In this form the equation is very easy to solve.
But, our case is not so simple. We will deal with a more sophisticated approximation that 
leads to the Fokker-Planck equation~\cite{sve}.

To start with, we apply the Landau approximation which allows only soft 
scattering in the collision integral.
If we define $w(\textbf{p},\textbf{k})$ to be the rate of collisions which change the momentum of
the HQ from $\textbf{p}$ to $\textbf{p}-\textbf{k}$, we have
\be
\left[\frac{\partial f}{\partial t}\right]_{collisions}= \int d^{3}\textbf{k}[w(\textbf{p}+\textbf{k},\textbf{k})
f(\textbf{p}+\textbf{k})-w(\textbf{p},\textbf{k})f(\textbf{p})].
\ee
The second part of the integral corresponds to all those transitions that remove HQ from momentum $\textbf{p}$ to
$\textbf{p}-\textbf{k}$, and therefore, represents a net loss to the distribution function. Likewise, the first part
of the integral represents a net gain to the distribution function of HQ. With these, Eq.~\ref{BTE1} becomes:
\be
\frac{\partial f(\textbf{p},t)}{\partial t}=\int d^{3}\textbf{k}[w(\textbf{p}+\textbf{k},\textbf{k})
f(\textbf{p}+\textbf{k})-w(\textbf{p},\textbf{k})f(\textbf{p})]
\label{BTE2}
\ee
Eq.~\ref{BTE2} is a linear equation in f. We can simplify it by assuming the previously discussed
Landau approximation. Mathematically, this approximation amounts to assuming $w(\textbf{p},\textbf{k})$
to fall off rapidly to zero with $|\textbf{k}|$, i.e., transition probability function, $w(\textbf{p},\textbf{k})$
is sharply peaked around $|\textbf{k}|=0$. Therefore, if we expand the integrand in the right hand
side of Eq.~\ref{BTE2} in powers of $\textbf{k}$, we have
\be 
w(\textbf{p}+\textbf{k},\textbf{k})f(\textbf{p}+\textbf{k})\approx w(\textbf{p},\textbf{k})f(\textbf{p})
+\textbf{k}\cdot\frac{\partial}{\partial \textbf{p}}(wf)+\frac{1}{2}k_{i}k_{j}
\frac{\partial^{2}}{\partial p_{i}\partial p_{j}}(wf)
\ee
Retaining terms up to the second order only, we obtain Fokker-Planck equation \cite{sve}:
\be
\frac{\partial f}{\partial t}= \frac{\partial}{\partial p_{i}}\left[A_{i}(\textbf{p})f+\frac{\partial}{\partial p_{j}}[B_{ij}
(\textbf{p})f]\right]~~,
\label{landaukeq}
\ee
where the kernels are defined as the following:
\be
A_{i}= \int d^{3}\textbf{k}w(\textbf{p},\textbf{k})k_{i}~~,
\label{eqdrag}
\ee
and
\be
B_{ij}= \frac{1}{2} \int d^{3}\textbf{k}w(\textbf{p},\textbf{k})k_{i}k_{j}.
\label{eqdiff}
\ee
In the present formalism, we considered the elastic scattering of the HQ with the gluon, light quark and
the corresponding anti-quarks. All these processes contribute to determine $w(\textbf{p},\textbf{k})$~\cite{sve}
and, in turn, the above defined kernels. 
Now, to explore the physical significance of the A and B coefficients, let us consider 
$A_{i}=p_i\gamma(p)$ and $B_{ij}=D(p)\delta_{ij}$, which assume very low $\textbf{p}$, i.e., the medium
of QGP is isotropic to the HQ. If, in this limit, we neglect all the derivatives of A and B coefficients
with momentum of HQ, Eq.~\ref{landaukeq} reduces to
\be
\frac{\partial f}{\partial t}=\gamma \frac{\partial}{\partial \textbf{p}}\cdot (\textbf{p}f)
+D\left[\frac{\partial}{\partial \textbf{p}}\right]^{2}f
\label{momindepfp}
\ee
The method of solution of this equation is elaborately discussed in \cite{sve}. 
Now, if we want to extend this formalism to the regime where the momentum of the HQ is no longer
small, i.e., the particular kinematic domain where HQ becomes relativistic, obviously, we would like 
to know how the transport coefficients drag($\gamma$) and diffusion(D) behave at high $\textbf{p}$ region.
In order to do so we extrapolate the concept of isotropy to the higher momentum of HQ in such a way that
only the first derivatives of the drag and diffusion coefficients are considered 
and the momentum dependence of A and B coefficients are encoded inside $\gamma$ and D. Therefore, the
Fokker-Planck equation, under this approximation, in Cartesian coordinate system becomes,~\cite{mbad}:
\bea
\frac{\partial f}{\partial t}&=&C_{1}(p_{x},p_{y},t)\frac{\partial^{2}f}{\partial p_{x}^{2}}~+C_{2}(p_{x},p_{y},t)
\frac{\partial^{2}f}{\partial p_{y}^{2}}\nn\\&+&~C_{3}(p_{x},p_{y},t)\frac{\partial f}{\partial p_{x}}~+C_{4}(p_{x},p_{y},t)
\frac{\partial f}{\partial p_{y}}\nn\\&+&~C_{5}(p_{x},p_{y},t)f~+C_{6}(p_{x},p_{y},t)
\label{fpeqcartesian}~~
\eea
where,
\bea
C_{1}&=& D\\C_{2}&=& D\\
C_{3}&=& \gamma ~p_{x}~+2~\frac{\partial D}{\partial p_{T}}~
\frac{p_{x}}{p_{T}}\\C_{4}&=& \gamma ~p_{y}~+2~\frac{\partial D}{\partial p_{T}}~\frac{p_{y}}{p_{T}}
\\C_{5}&=& 2~\gamma ~+\frac{\partial \gamma }{\partial p_{T}}~\frac{p_{x}^{2}}{p_{T}}~+
\frac{\partial \gamma }{\partial p_{T}}~\frac{p_{y}^{2}}{p_{T}}\\C_{6}&=& 0~~.
\eea
where the momentum, $\textbf{p}=(\textbf{p}_T,p_z)=(p_x,p_y,p_z)$. 
We numerically solve Eq.~\ref{fpeqcartesian} ~\cite{antia} with the boundary conditions:
$f(p_x,p_y,t)\rightarrow 0$ for $p_x$,$p_y\rightarrow \infty$ and  the initial (at time $t=\tau_i$)
momentum distribution of charm and bottom quarks are taken from MNR code~\cite{MNR}.
It is evident from Eq.~\ref{fpeqcartesian} that with the momentum dependent 
transport coefficients the FP equation becomes complicated. 

It is possible to write down the solution of the FP equation in closed analytical 
form~\cite{rapp} in the special case of momentum independent drag and diffusion coefficients.
To find out the solution $f$ for momentum independent drag and diffusion coefficients
we can consider the one-dimensional version of Eq. {\ref{momindepfp}}, 
\bea
\frac{\partial f}{\partial t}=\gamma \frac{\partial}{\partial}(pf)
+D \frac{\partial^2 f}{\partial p^2}~,
\label{momindepfp1d}
\eea
also called the Rayleigh's equation. For
the initial condition $f(p,~t=0)=\delta(p-p_0)$, the solution of
$f$ is
\bea
f(p,t)&=&\left(\frac{\gamma}{2\pi D}(1-e^{-2\gamma t})\right)^{-1/2}\nn\\
&&\times exp{\left(-\frac{\gamma}{2D}\frac{(p-p_0 e^{-\gamma t})^2}{1-e^{-2\gamma t}}\right)}
\label{fmomindep}
\eea

However, Eq. \ref{fmomindep} is solution for a very simplified scenario and we will
study the momentum dependence of drag and diffusion coefficients and their effects on 
$R_{AA}$ in the sections to come.
\section{Drag Coefficients:}
\subsection{Elastic processes:}
We need to evaluate the drag coefficient as a function of temperature and momentum of HQ. 
The expression to evaluate collisional drag can be written as~\cite{sve}:
\bea
\gamma_{coll}&=& \frac{1}{2E_{p}} \int \frac{d^{3}\textbf{q}}{(2\pi)^{3}2E_{q}}~\int \frac{d^{3}\textbf{q}'}{(2\pi)^{3}2E_{q'}}\nn\\
&\times&\int \frac{d^{3}\textbf{p}'}{(2\pi)^{3}2E_{p'}}~\frac{1}{\gamma_{Q}}\sum |M|^{2}
\nn\\&\times&(2\pi)^{4}\delta^{4}(p+q-p'-q')f'(\textbf{q})[1-\frac{\textbf{p}.\textbf{p}'}{p^{2}}]
\label{drag}
\eea
where $\textbf{p}'=\textbf{p}-\textbf{k}$ and $\textbf{q}'=\textbf{q}+\textbf{k}$.
The scattering matrix elements are given explicitly in \cite{combridge} where the 
$t$ channel divergence occurring due to very soft gluon exchange has
been shielded by replacing $t$ in the denominator of matrix elements by $t-m_D^2$ in
an ad hoc manner, where $t$ is the Mandelstam variable and $m_D$ is the thermal mass of gluon. 
However, the same problem can be approached from hard thermal loop (HTL) 
perturbation theory. The gluon propagator for $t$ channel diagram is then replaced 
by HTL propagator. But, calculation of even the elastic matrix elements for HQs scattering 
with light quarks and gluons in this approach is very lengthy and radiative matrix elements
are even clumsier. However, an outline of calculating collisional 
drag and diffusion coefficients in HTL perturbation theory approach is given for 
the interested readers in the Appendix. Here we  proceed with
the conventional process of shielding $t$ channel divergence with the gluon thermal mass, $m_D$. 

\begin{figure}[h]
\begin{center}
\includegraphics[scale=0.43]{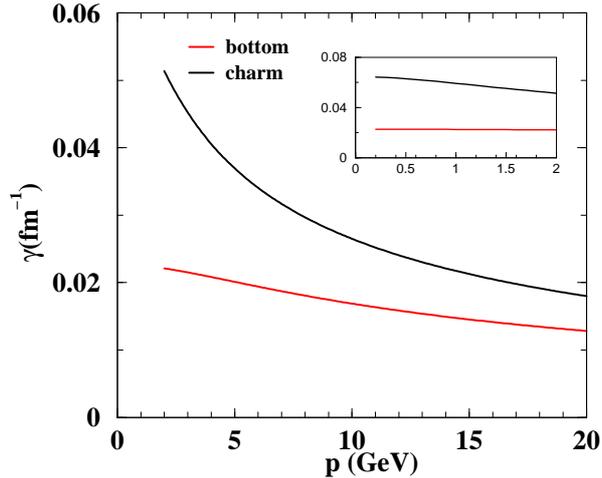}
\caption{Drag coefficients of charm and bottom with their momentum at T=300 MeV assuming running 
strong coupling,  $\alpha_{s}(T)$ and temperature dependent
Debye screening mass, $m_{D}(T)$ for gluon, quark, and antiquark scattering.}
\label{FIG2}
\end{center}
\end{figure}

The integrations in Eq.~\ref{drag} has been performed using the 
standard techniques~\cite{das,san,dams,sve}. Results of the present calculation of drag 
coefficients are plotted with respect to momenta of charm and bottom (Fig.~\ref{FIG2}).
We can observe that the momentum dependence of $\gamma$ is non-negligible for the shown
momentum range. The value of $\gamma$ for charm due to collisional 
processes at $p=5$ GeV is about 0.036 fm$^{-1}$ which reduces to a value of
0.018 fm$^{-1}$ at $p=10$ GeV 
(can be compared with ~\cite{dkspalmunshi,grecorappepj,gossiauxetal,herappfries} for example).
In the inset, the drag coefficients of HQ's due to elastic collision are plotted in the lower 
momentum region, where the drag remains more or less constant with $\textbf{p}$. Therefore,
it is clear that had we taken the value of drag at low momentum and extrapolated 
that value to higher momentum, the final result would have been overestimated.
The diffusion coefficient of HQ can be evaluated from Einstein relation $D=\gamma MT$.
Later, it will be seen that the momentum dependence of transport coefficients of a high energy 
HQ will have considerable effects on the nuclear suppression factor, $R_{AA}$ of HQ.

However, there are also radiative processes taking place in the medium. The radiative 
drag can be obtained in the present formalism by computing the radiative energy loss
of a HQ passing through QGP. In the next section we will discuss the methods and intricacies
of calculating the transport coefficients in radiative domain.
 

\subsection{Radiative Processes:}
It is already known that the two mechanisms of energy loss of heavy quarks 
are collisional and radiative energy loss. Though at low transverse momentum ($p_T$) 
region the collisional and radiative losses of HQs are comparable (see Fig. \ref{p1dragfig1}), the radiative one
tends to dominate with increasing momentum. So it is worthwhile,
after discussing about 
evaluation of collisional transport coefficients (drag, diffusion)
within the ambit of $T=0$ perturbative QCD (pQCD) \cite{sve} in the previous section, to 
contemplate on radiative processes and to inspect how pQCD approach can be utilized in 
finding out radiative transport coefficients.

It is well-known from QED that for high energies 
radiative energy loss becomes dominant\cite{leo}. Also the Hard Thermal Loop
calculations in context of QCD show that the radiative energy loss contributes to same order
of strong coupling as that of collisional loss~\cite{thomaprdrapid}. 
The above discussions tempt one to infer that the observed large suppression of
heavy quarks at RHIC is predominantly due to bremsstrahlung processes~\cite{dams,mbad}, 
at least at high $p_T$; but, as already said, the comparable values of energy losses 
in low momentum region leaves an ample room for ambiguity in this
statement particularly at `not-so-high' $p_T(\sim2~GeV)$ region. 

\begin{figure}[h]
\begin{center}
\includegraphics[scale=0.43]{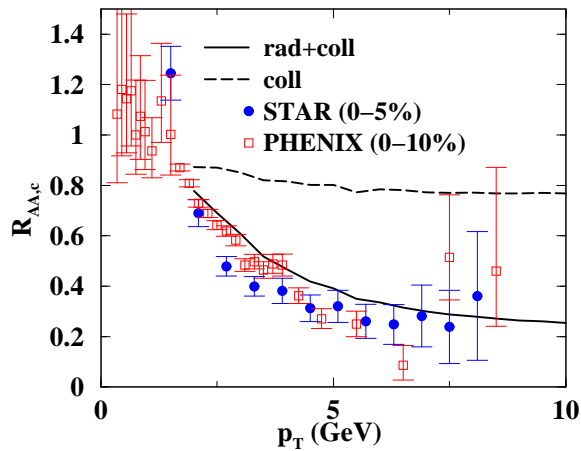}
\caption{Suppression of transverse momentum of charm 
quarks in QGP as a function $p_{T}$ 
}
\label{RAA-radcollC}
\end{center}
\end{figure}

Theoretical estimate of nuclear
suppression factor ($R_{AA}$) for charm quarks in \cite{mbad} shows $\sim$4 times 
more suppression due to inclusion of radiation (Fig. \ref{RAA-radcollC}) with same initial 
conditions. Inclusion of radiative processes leads to a good description of charm $R_{AA}$
at RHIC energies.
\noi At LHC energy (2.76 TeV/A) the collisional energy loss due to hard (momentum transfer $> 2 GeV$)
collisions could be about one-third 
of the total~\cite{younus}. The rest may be attributed to radiative loss. 

From the discussion on collisional transport coefficients we know that (collisional) drag
is given by~\cite{sve}
\bea
\gamma_{coll}=-\frac{1}{p}\left(\frac{dE}{dx}\right)_{coll},
\eea
where $p$ is the momentum of the probe. We employ
a similar argument and relate radiative energy loss ($(dE/dx)_{rad}$) to
inelastic drag ($\gamma_{rad}$) in the same way. The effective drag obtained is a summation
of collisional and radiative parts. As energy loss is related with the transport properties like 
drag offered by the medium, we must concentrate on more and more accurate
determination of radiative energy loss which will enable us to understand the properties 
of QGP. 

The radiative energy loss in general has been studied in \cite{gptw,wgp,gw,baieretal,glv}
incorporating Landau-Pomeranchuk-Migdal (LPM) effect~\cite{lpm}, to be discussed later, due to multiple scattering. 
\cite{gossiaux} treats the problem of radiative energy loss of HQs by building 
a model in scalar QCD approach. Attempts have been taken ~\cite{dams,mbad,mustafa} 
to incorporate the formalism of ~\cite{gptw}, the Gyulassy-Wang potential model(GWPM), to compute 
the radiative energy loss of heavy quarks traveling through QGP. In stead of going to
the energy loss in GWPM approach directly, we may try to acquire some familiarity with the model
and the notations used for the description of it.

\subsubsection{Gyulassy Wang Potential Model (GWPM) and radiation spectrum:}
\begin{figure}[h]      
\includegraphics[scale=0.3]{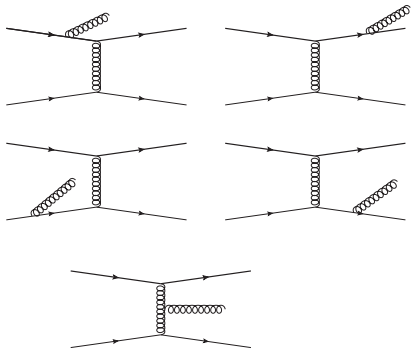}                                 
\caption{Feynman diagrams for gluon radiation from a
single quark-quark ($qq^{\prime}$) scattering }
\label{fig1}                                                               
\end{figure}

To analyze the multiple scattering and the induced gluon radiation in GWPM,
certain simplifications have to be made. For example, \cite{wgp}
assumes static interaction between propagating parton and bath particles.
This interaction is modelled by a static Debye screening potential. Now,
it is possible to approximate the effective average random colour field
produced by the bath particles by a potential provided the distance between two
successive scatterers is large compared to colour screening length ($\mu^{-1}$). 
The screened potential is given by:
\bea
V_{AA'}^a(\vec{q})&=&A_{AA'}^a(\vec{q})e^{-i\vec{q}.\vec{x}}\nn\\
&=&gT_{AA'}^a\frac{e^{-i\vec{q}.\vec{x}}}{\vec{q}^2+\mu^2},
\label{eq1}
\eea
where $\mu$ is the colour screening mass, $T^a$ are the generators corresponding
to the representation of target partons at $\vec{x}$ transferring (three) momentum 
$\vec{q}$ and $g$ is the coupling.
The Feynman diagrams contributing to induced gluon radiation from a
single quark-quark scattering is given in Fig.~\ref{fig1}.

The calculation of Feynman diagrams are done in light-cone co-ordinates 
\cite{lightcone} where light-cone representations of four-vectors are done
in the following way,
\bea
p&=&(p^+,p^-,p_{\perp})\nn\\
&=&(p_0+p_3,p_0-p_3,p_{\perp})
\label{eq2}
\eea
In the potential model one can neglect radiation from the target lines
(external lines appearing at the bottom of gluon propagator in Fig.~\ref{fig1}) 
provided one decides to work in light cone gauge, $A^+=0$, for emitted gluon fields 
(For details see~\cite{rthomas}). For radiation in the mid-rapidity region the celebrated Gunion-Bertsch distribution
formula (GB formula) for soft gluon radiation ~\cite{gb} can be reproduced from this approach.

We now discuss the GB formula and recent attempts to generalize \footnote{Interested readers
are referred to the Appendix for other such very recent endeavors.}it
because we will see that this formula and/or its generalizations  will play a
significant role in finding out the energy loss due to gluon emission of quarks.
The early attempts of generalizing GB formula~\cite{DA,GR,bmad} consider the general gg $\rightarrow$ ggg
matrix element elegantly written by ~\cite{berendes} and factoring out the elastic gg $\rightarrow$ gg
scattering amplitude to obtain the distribution of emitted gluon 
($\sim|M_{\mathrm{gg} \rightarrow \mathrm{ggg}}|^2/|M_{\mathrm{gg} \rightarrow \mathrm{gg}}|^2$). 
We can write the following form of gluon distribution~\cite{bmad},
(for details see Appendix of ~\cite{bmad})
\bea
\frac{dn_\mathrm{g}}{d^2 k_{\perp}d\eta}&=& \left[ \frac{dn_{\mathrm{g}}}{d^2 k_{\perp}d\eta}\right]_ {GB}
[\left( 1+\frac{t}{2s}+\frac{5t^2}{2s^2}-\frac{t^3}{s^3} \right)\nn\\
&-&\left( \frac{3}{2\sqrt{s}}+\frac{4t}{s\sqrt{s}}-\frac{3t^2}{2s^2\sqrt{s}} \right) k_{\bot} \nn\\
&+& \left( \frac{5}{2s}+\frac{t}{2s^2}+\frac{5t^2}{s^3} \right)k_{\bot}^2 ],
\label{spectrum}
\eea
where $\eta$ is the rapidity of the radiated gluon, the subscript GB has been used to indicate the
gluon spectrum obtained using the approximation considered in~\cite{gb} (see
also~\cite{wong})
which is generally given by,
\be
\left[\frac{dn_{\mathrm{g}}}{d^2 k_{\perp}d\eta}\right]_ {GB}=
\frac{C_A \alpha_s}{\pi^2}\frac{q_{\perp}^2}{k_{\perp}^2[(\vec{k}_{\perp}-\vec{q}_{\perp})^2
+m_D^2]},
\label{gbspectrum}
\ee
where $m_ D=\sqrt{\frac{2\pi}{3}\alpha_s(T)\left(C_A+\frac{N_F}{2}\right)}~T$,
is the thermal mass of the gluon~\cite{Bellac,kap}, $N_F$ is the number of flavours contributing
in the gluon self-energy loop,
$C_A=3$ is the Casimir invariant for the SU(3) adjoint representation,
$\alpha_s$ is the temperature-dependent strong coupling~\cite{KACZ}, $k_{\perp}$ is
the transverse momentum of the emitted gluon 
and $q_{\perp}$ is the transverse momentum transfer. In general, one introduces
the thermal mass in the denominator of Eq.~\ref{gbspectrum}
to shield the divergence arising from collinearity
({\it i.e.} when $\vec{k}_{\perp}=\vec{q}_{\perp}$) of emitted gluon. 
But, for the present case Eq. \ref{spectrum} is written under the assumption 
that $\vec{q}_{\perp}>>\vec{k}_{\perp}$; and hence, there is no need to write
the $m_D^2$ factor in GB spectrum.

Now, we must consider multiple scattering encountered by incoming particles, too. The many body
effect due to presence of medium results in interference of scattering amplitudes. This 
interference effect is called LPM effect~\cite{lpm}. LPM effect is discussed 
in QED domain in \cite{klein}. LPM suppression can be understood in a qualitative manner as
an interplay between two time scales, the formation time ($\tau_f$) and the scattering time ($\tau_{sc}$). 
$\tau_F$ is the time needed for the emission of the induced gluon. Actually, $\tau_f$
determines the time span after which a radiation can be separately identified from the parent
parton from which the radiation is being given off. Now, a collision
just before formation of the gluon results in suppression of the radiation. This destructive
interference is called LPM effect and in the next section we will see that it
puts a constraint on the phase space of the emitted gluon. 
If $k_0$ is the energy of the emitted (soft) gluon and $k_{\perp}$
is its transverse momentum, then formation time $\tau_f\sim\frac{2k_0}{k_{\perp}^2}$~\cite{wgp}. When this $\tau_f$
is much less than collision time {\it i.e.}, separation between two scattering centres, $L$, then
the intensity of radiation is additive and {\it Bethe-Heitler(BH) limit} is reached. In the {\it factorization limit}
, $L<<\tau_f$, the interference in radiation amplitude takes place and LPM effect dominates. 

After this brief discussion on possible effects of multiple scattering, 
emitted gluon distribution due to multiple scattering in terms of that due to single scattering 
can be written as~\cite{wgp}:
\be
\frac{dn_{\mathrm{g}}^{(m)}}{d^2 k_{\perp}d\eta}=C_m(k) \frac{dn_{\mathrm{g}}^{(1)}}{d^2 k_{\perp}d\eta},
\ee
where `$m$' stands for multiple scattering and `$1$' stands for single scattering. $C_m$ is called 
radiation formation factor characterizing the interference pattern due to multiple 
scattering. Naturally, in the BH limit, $C_m\approx m$, {\it i.e.} the scatterings
add up to give the resultant intensity with no interference pattern. On the other hand,
the factorization limit gives~\cite{wgp,gw},
\bea
C_m(k)&\approx& \frac{8}{9}[1-(-1/8)^m]~~~for~quarks\nn\\
&\approx&2(1-1/2^m)~~~~~~~~for~gluons          
\label{cm}
\eea
Eq.~\ref{cm} shows that the interference effect due to many multiple scatterings 
for quarks leaves corresponding radiation spectrum a factor of $\sim 8/9$ of that due to
single scattering. It can also be checked that the gluon intensity radiated by gluon jet 
is $9/4$ times higher than that radiated by quark jets in multiple scattering. Thus,
the LPM effect in QCD depends on colour representation due to non-abelian 
nature of the problem under discussion. 

\subsubsection{Energy Loss of light particles in GWPM:}
\begin{figure}[h]                                                            
\begin{center}                                                               
\includegraphics[scale=0.40]{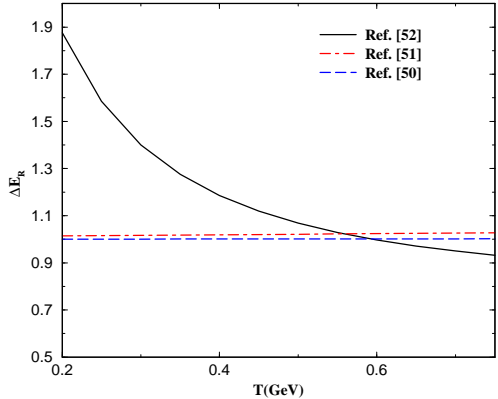}                                 
\end{center}                                                                 
\caption{(Color online) Temperature variation of $\Delta E_R$, 
{\it i.e.} $\Delta E$ normalized by 
the corresponding value obtained from GB approximation,
of a 15 GeV gluon 
due to traversal of 4 fm 
in a gluonic heat bath. Solid (dashed) line indicates result for
the gluon spectrum obtained in  \cite{bmad} (\cite{DA})). 
The dot-dashed line stands for the results 
for the gluon spectrum of ~\cite{GR}.}                                                          
\label {fig3}                                                               
\end{figure}

The energy loss in potential model can be evaluated by integrating
over the transverse momentum and rapidity of emitted gluon. The phase
space is, of course, constrained by LPM effect due to multiple scattering.
The multiple scattering is implemented through
the differential change of the factor $C_m$ with number of scattering m, $\frac{dC_m}{dm}$,
which can be approximated as a $\theta$-function~\cite{wgp}. Also, the emitted gluon must have
energy less than that of the parent parton. So the energy loss formula implementing the energy 
constraint gives:
\bea
\Delta E_{rad}&=&\frac{E_{m+1}-E_m}{m}\nn\\
&=&\int d^2 k_{\perp} d\eta \frac{dn_{\mathrm{g}}}{d^2 k_{\perp}d\eta} k_0 \frac{C_{m+1}-C_m}{m}
\theta(E-k_{\perp}cosh\eta)\nn\\
&=&\int d^2 k_{\perp} d\eta \frac{dn_{\mathrm{g}}}{d^2 k_{\perp}d\eta} k_0 \frac{dC_m}{dm}
\theta(E-k_{\perp}cosh\eta)\nn\\
&\sim&\int d^2 k_{\perp} d\eta \frac{dn_{\mathrm{g}}}{d^2 k_{\perp}d\eta} k_0 \theta(\tau_{sc} -\tau_F)
\theta(E-k_{\perp}cosh\eta)
\label{eloss}
\eea
where $E$ is the energy of the parent parton and $k_0=k_{\perp}cosh\eta$, 
$E_{m+1}$ denotes energy
loss in $(m+1)^{th}$ collision and $E_m$ is that in $m^{th}$ collision. The first $\theta$-function involving
scattering time $\tau_{sc}$ \cite{thoma}
gives a lower limit of $k_{\perp}$, 
{\it i.e.} $k_{\perp}>\Lambda cosh\eta$. The second $\theta$-function yields,
$k_{\perp}<E/cosh\eta$. Utilizing Eq.~\ref{eloss}, ~\cite{bmad} compares the 
energy loss (Fig.~\ref{fig3}) of a gluon jet passing through gluonic plasma 
obtained by generalized GB formulae given in  \cite{DA,GR,bmad}. We will see 
how Eq. \ref{eloss} can be employed to find out HQ energy loss in the next section.

\subsubsection{Energy Loss of HQs and radiative drag:}
While calculating the radiative energy loss
of HQs ~\cite{gossiaux} takes the GB distribution for small and moderate 
$x$ and for mass $m\neq 0$. The distribution, so obtained, can be shown to yield GB distribution
when $x<<1$ and when $m=0$. Whereas,~\cite{dams,mbad} use the original GB distribution formula
weighted by a radiative suppression factor, called `Dead-cone' factor, originating due to mass
of HQs.  Since dead-cone factor plays an important role
in radiative energy loss of heavy quarks, we will pause here to spend some words on 
Dead-cone effect in QCD \cite{khar} and its generalizations.

The dead cone suppression obtained in~\cite{khar} actually has an analogy with radiated power distribution 
of a non-relativistic, accelerating charge particle. The average power radiated per unit solid angle is 
given by~\cite{jackson}:
\be
\frac{dP}{d\Omega}\propto |\dot{\vec{\beta}}|^2 sin^2 \theta,
\label{power}
\ee
where $\theta$ is the angle between acceleration $\dot{\vec{\beta}}$ of the particle and the direction of 
propagation of radiation, $\vec{n}$. This is a simple $sin^2 \theta$ behaviour showing no radiation 
at $\theta=0~(\rm{or~n}\pi, n\in Z)$. It can be shown ~\cite{kmpaul} that the behaviour of Eq.~\ref{power}
is, indeed, similar to what one  gets for conventional dead cone~\cite{khar}.
Qualitatively speaking, it is hard to cause a deceleration of a high energy heavy quarks
along their directions of motion; and this is why the bremsstrahlung radiation is suppressed 
along this direction (see Fig. \ref{figkmpaul}). Indeed, the distribution of radiated gluons emanating
from HQs are shown to be related to those from light quarks (LQs) by the following
formula~\cite{khar} (for small radiation angle $\theta$):
\bea
dP_{HQ}=\left(1+\frac{\theta_0^2}{\theta^2}\right)^{-2} dP_{LQ},
\label{khardead}
\eea
where $\theta_0=m/E$; $m$ is the mass and $E$ is the energy of heavy quark. It is worth noting 
that when $E\rightarrow\infty$, the radiation from HQs converges  with that of LQs. 
However, there are some very recent developments in generalizing the dead-cone effect 
either from $Qq\rightarrow Qq$g matrix element or assuming effects of off-shellness
of quarks which, after being produced, take some time to become on-shell. The details
may be seen in the Appendix.

\begin{figure}[h]
\begin{center}
\includegraphics[scale=0.43]{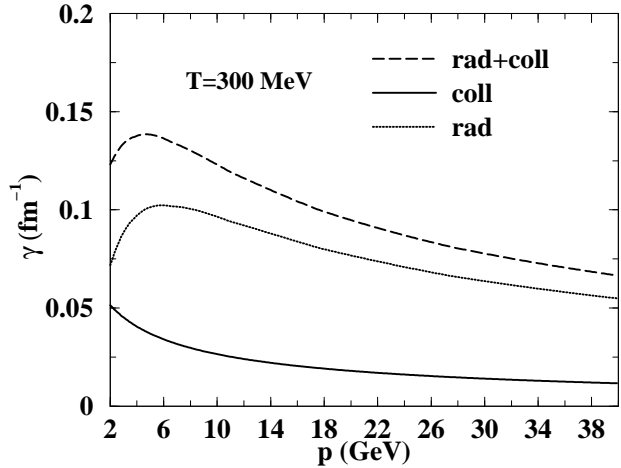}
\caption{Drag coefficients of charm assuming running 
strong coupling,  $\alpha_{s}(T)$ and temperature dependent
Debye screening mass, $m_{D}(T)$ due to its interaction
with thermal gluons, quarks, and antiquarks.}
\label{p1dragfig1}
\end{center}
\end{figure}

The radiative transport coefficients like drag, diffusion have been evaluated
in~\cite{dams,mbad} employing potential approach. 
The radiative drag coefficient can be obtained by finding out the radiative  energy loss of HQs 
passing through medium. The GB spectrum for emitted gluon,
weighted by the dead-cone factor and the energy of gluon $k_0$ ($=k_{\perp}cosh\eta$),
is integrated over the transverse momentum ($k_{\perp}$)
and rapidity ($\eta$) of the emitted gluon. As already stated,
the lower and upper limits of $k_{\perp}$ has been obtained from the 
$\theta$-functions of Eq.~\ref{eloss}. 
The average energy loss per collision, $\Delta E_{rad}$,
can be written with the help of Eq. \ref{eloss},
\bea
&=& \langle n_{g}k_{0}\rangle= \int d\eta~d^{2}k_{\bot}~\frac{dn_{\mathrm{g}}}{d\eta d^{2}k_{\bot}}\nn\\
&\times&k_{0}~\Theta (\tau_{sc} -\tau_{F})~\Theta (E-k_{\bot}\cosh\eta)F_{DC}^2~~,
\label{avgenergy}
\eea
where the formation time of the emitted gluon~\cite{wgp}, 
$\tau_{F}=(C_{A}/2C_{2})~2\cosh\eta/k_{\bot}$, and $C_{A}/2C_{2}=N^{2}/(N^{2}-1)$
for quarks with $C_{2}=C_{F}=4/3$. Dead-cone factor of Eq.~\ref{khardead} ($dP_{HQ}/dP_{LQ}$)
can be written in the following way, provided one replaces 
$k_{\perp}/k_0\sim sin\theta \sim \theta$ for small $\theta$,
\be
F_{DC}^2= \left(\frac{k_{\bot}^{2}}{k_{0}^{2}\theta_{0}^{2}+k_{\bot}^{2}}\right)^2~~,
\ee
$dE/dx$ can be obtained if one multiplies $\Delta E_{rad}$ with $\Lambda$, which can be obtained 
from \cite{thoma}. Drag ($\gamma$) and diffusion ($D=\gamma m T$) coefficients in \cite{mbad}
are functions of momentum as well as temperature. The effective drag can be
obtained by adding collisional and radiative drags, $\gamma_{eff}=\gamma_{coll}+\gamma_{rad}$.
The momentum dependence of the drag coefficient of the charm quark propagating
through the QGP is displayed in Fig. \ref{p1dragfig1} for $T=300$ MeV. For $p_T\sim$ 2 GeV of charm
quarks the collisional and radiative contributions tend to merge with each other. However, 
it is interesting to note the dominance of radiative drag in Fig. \ref{p1dragfig1} 
over its collisional counterpart for higher momentum.

\section{Space-Time Evolution}
Once we know the total drag ($\gamma_{eff}$) and diffusion ($D_{eff}$) coefficients, we need the initial
conditions for both the HQ and the QGP background which is also evolving with time following,
\be
\partial_{\mu} T^{\mu \nu}=0
\label{enermomenconservation}
\ee
where, $T^{\mu\nu}=(\epsilon+P)u^\mu u^\nu-g^{\mu\nu}P$, 
is the energy momentum tensor for ideal fluid, 
$\epsilon$ is the energy density, $P$ is the pressure
and $u^\mu$ is the hydrodynamic four velocity, and $g^{\mu\nu}$
is the metric tensor. We will solve this equation for longitudinal expansion 
assuming boost invariance along the $z$ 
direction~\cite{jdbjorken}. It is expected that the central rapidity
region of the system formed after nuclear collisions
at RHIC and LHC energy is almost net baryon free. Therefore,
the equation governing the conservation of net
baryon number need not be considered here.
Under this circumstance Eq.~\ref{enermomenconservation}
reduces to:
\be
\frac{\partial\epsilon}{\partial\tau}+\frac{\epsilon + P}{\tau}=0
\label{hyy1}
\ee
To solve Eq.~\ref{hyy1}, we use the relation 
$P=c_s^2\epsilon$, where $c_s$ is the velocity of sound and $\tau$ is time.
With this we arrive at the relation,
\be
\epsilon \tau^{1+c_s^2}=C
\ee 
where $C$ is a constant.
For the ideal equation of state, $c_s^2=\frac{1}{\sqrt{3}}$, the above Eq. reads
$\epsilon \tau^{\frac{4}{3}}=C$. In terms of temperature this relation 
can be written as $\tau T^3=C$. 

The total amount of energy dissipated by a  heavy quark in the expanding QGP
depends on the path length it traverses.
Each parton traverses different path length
which depends on the  geometry of the system and on the point 
where its is created.
The probability that a parton is produced at a point $(r,\phi)$
in the plasma depends on the number of binary collisions 
at that point which can be taken as~\cite{turbide}:
\be
P(r,\phi)=\frac{2}{\pi R^2}(1-\frac{r^2}{R^2})\theta(R-r)
\label{prphi}
\ee
where $R$ is the nuclear radius. It should be mentioned here
that the expression in Eq.~(\ref{prphi}) is an approximation for the
collisions with zero impact parameter. 
A very high energy parton created at $(r,\phi)$ in the transverse plane
propagates a distance $L=\sqrt{R^2-r^2sin^2\phi}-rcos\phi$ 
in the medium provided its direction remains unaltered. In the present work we use the following
equation for the geometric average of the integral
involving drag coefficient 
\be
\Gamma=\frac{\int rdr d\phi P(r,\phi) \int^{L/v}d\tau\gamma(\tau)}
{\int rdr d\phi P(r,\phi)},
\label{cgama}
\ee
where $v$ is the velocity of the propagating partons. 
Similar averaging has been performed for the diffusion coefficient, too.
For a static system the temperature dependence of the drag and
diffusion coefficients of the heavy quarks enter via the
thermal distributions of light quarks and gluons through
which it is propagating. However, in the present scenario
the variation of temperature with time is governed by
the equation of state or velocity of sound
of the thermalized system undergoing hydrodynamic
expansion. In such a scenario the quantities like $\Gamma$ (Eq.~\ref{cgama})
and hence the HQ suppression becomes sensitive to $c_s$.

\section{Initial Conditions and Nuclear Modification Factor}
In order to solve Eq.~\ref{fpeqcartesian}, the initial distribution functions, $f_{in}(p_{T},t)$ for charm and bottom
quarks have been supplied from the well known MNR code~\cite{MNR}. The ratio between the solution of Fokker-Planck
Equation at the transition temperature, $T_c=175$ MeV and the initial distribution function of HQ is the required 
nuclear modification factor, $R_{AA}$ of open HQ. But, in order to compare results from the present
formalism with experimental data from RHIC and LHC we need to have $R_{AA}$ of non-photonic single electron 
originating from the decays of D and B mesons.

Therefore, the hadronization of charm and bottom quarks
to $D$ and $B$ mesons respectively are done by using
Peterson fragmentation function~\cite{peterson}:
\be
f(z) \propto \frac{1}{\lbrack z \lbrack z- \frac{1}{z}- \frac{\epsilon_Q}{1-z} \rbrack^2 \rbrack}
\label{petersonfrag}
\ee
where, z is the fraction of momentum carried by the hadrons, $\epsilon_Q$ is 0.05 for
charm and $(m_c/m_b)^2 \epsilon_Q$ for bottom where $m_c(m_b)$ is the mass of charm(bottom).
One may use different kinds of available fragmentation functions, but the final result
will not be sensitive to the choice of $f(z)$. In point of fact, describing hadronization has always
been a formidable task because the hadron bound states are non-perturbative in nature.
The fragmentation scheme is a popular method to deal with hadronization.
For more relevant information interested readers are referred to \cite{coalescencegreco}, 
which discusses coalescence model, a hadronization formulation based on recombination 
of heavy flavours. 

Both the final solution of FP Equation and the initial distribution of HQ are convoluted with
the above fragmentation function (Eq.~\ref{petersonfrag}) and their ratio will give $R_{AA}$ of
D and B mesons. The final and initial distribution functions are obtained for the single electrons
originated from the decays of D and B mesons and the final nuclear modification factor is:
\be
R_{AA}^{D(B)\rightarrow e}=\frac{f^{D(B)\rightarrow e}(p_T,T_c)}{f^{D(B)\rightarrow e}(p_T,T_i)}
\label{raa}
\ee
Now, once we know how to determine $R_{AA}$, we will compare our results with the RHIC and LHC data. In
the process of doing so we will study the effects of the equation of state on the
nuclear suppression of heavy flavours in quark gluon plasma and estimate the initial entropy density 
of the QGP formed at the RHIC. For this purpose, the experimental data on the charged particle multiplicity and the nuclear 
suppression of single electron spectra originating from the semi-leptonic decays of D and B mesons have 
been employed. We have used inputs from lattice QCD (LQCD) to  minimize the model dependence of the results.

The initial entropy density and the thermalization time ($\tau_i$) for the QGP can be constrained 
to the measured (final) multiplicity by the following relation~\cite{hwa} which is boost invariant:
\be
s_i\tau_i=\kappa\frac{1}{A_\perp}\frac{dN}{dy}
\label{dndy}
\ee
where $A_\perp$ is the transverse area of the system which can be
determined from the collision geometry and
$\kappa$ is a known constant (=3.7 for massless Bosons).

\begin{figure}[h]
\begin{center}
\includegraphics[scale=0.43]{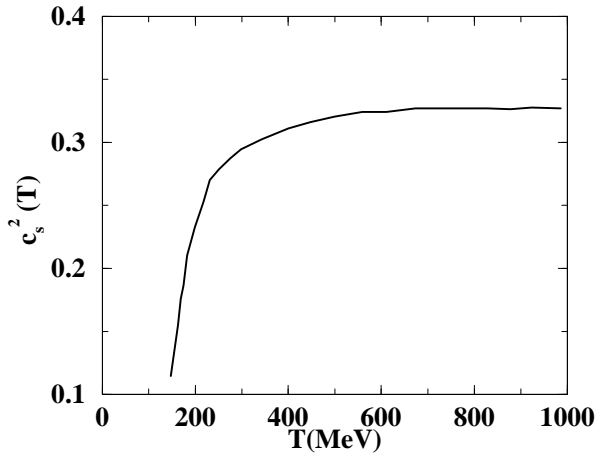}
\caption{Velocity of sound squared as a function temperature~\cite{borsanyi}}
\label{FIG3}
\end{center}
\end{figure}

Equation of state (EoS), which is taken as $P=c_s^2\epsilon$ for almost baryon free QGP
expected at RHIC energy, also has its own role to play in the space-time evolution of QGP.
This sound velocity squared appearing in the EoS shows a significant variation with
temperature in LQCD calculations (fig.\ref{FIG3}).
\begin{figure}
\begin{center}
\includegraphics[scale=0.43]{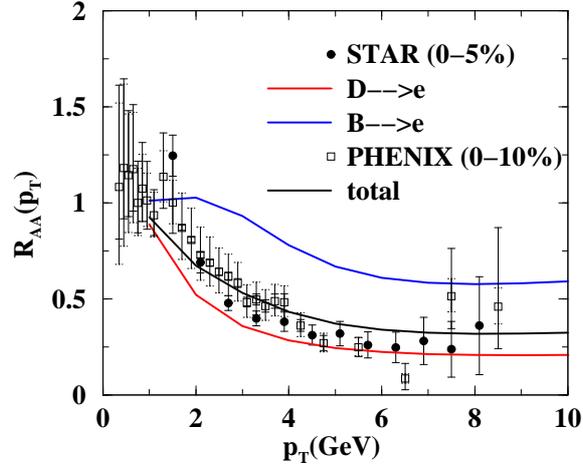}
\caption{(colour online) Variation of $R_{AA}$  with
$p_T$ for the space time evolution with initial condition $T_i=250$ MeV 
and $\tau_i=0.84$ fm/c and the EoS which includes the variation of $c_s$
with $T$.}
\label{FIG4}
\end{center}
\end{figure}
\begin{figure}
\begin{center}
\includegraphics[scale=0.43]{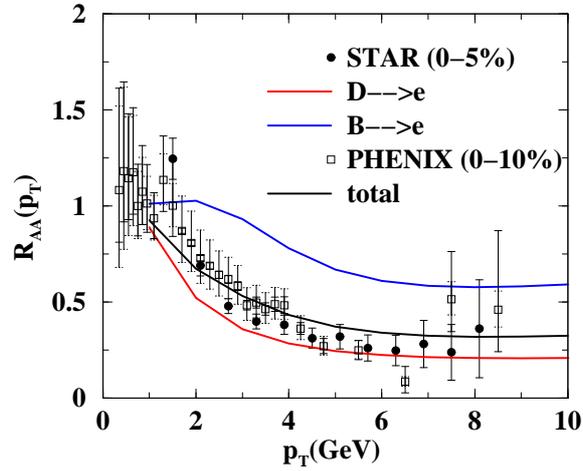}
\caption{(colour online) Variation of $R_{AA}$ with $p_T$ for 
for $c_s^2=1/3$ and $T_i=300$ MeV.}
\label{FIG5}
\end{center}
\end{figure}
\begin{figure}
\begin{center}
\includegraphics[scale=0.43]{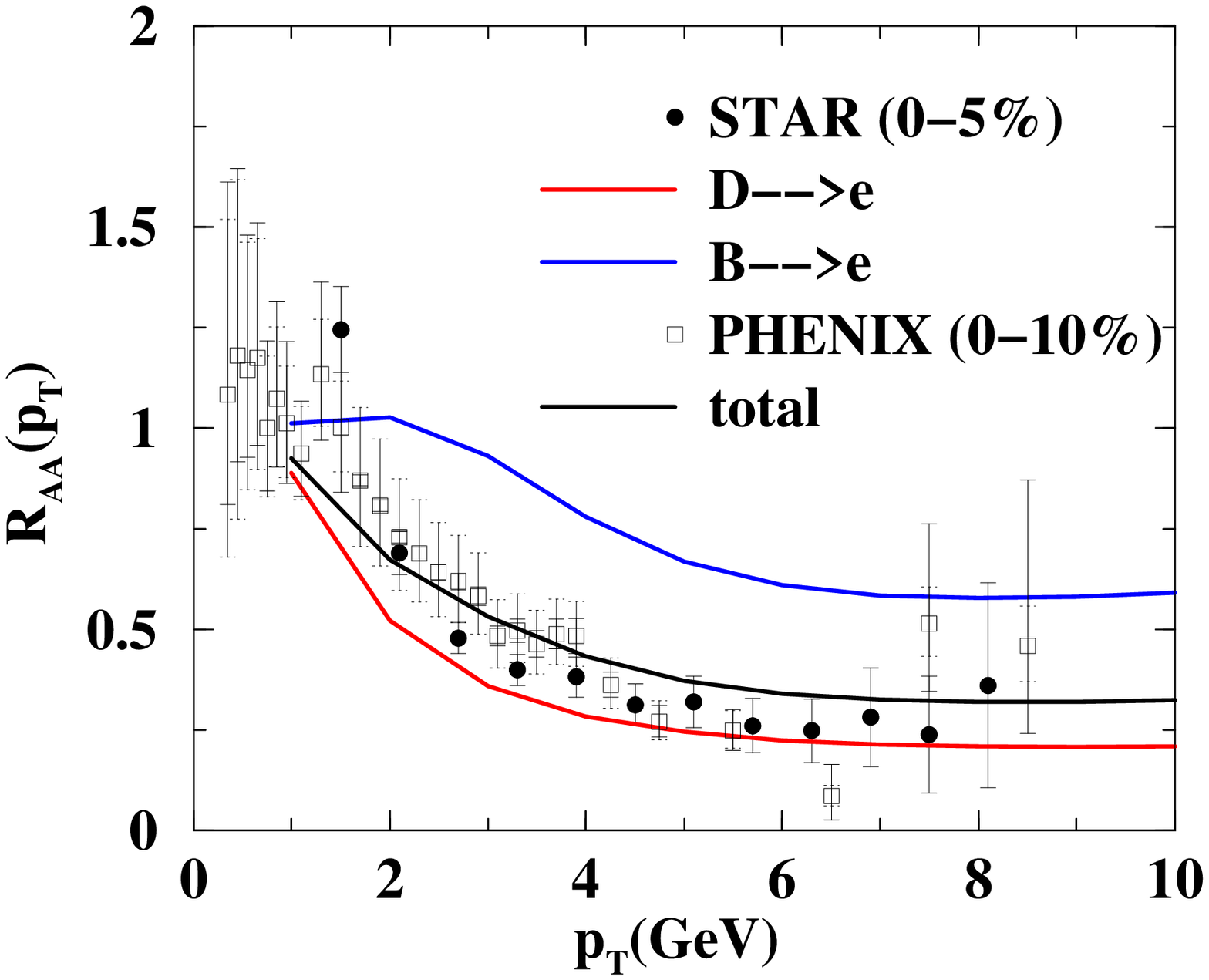}
\caption{(colour online) Variation of $R_{AA}$ with $p_T$ for $c_s^2=1/5$ 
and $T_i=210$ MeV.}
\label{FIG6}
\end{center}
\end{figure}

\begin{figure}[h]
\begin{center}
\includegraphics[scale=0.43]{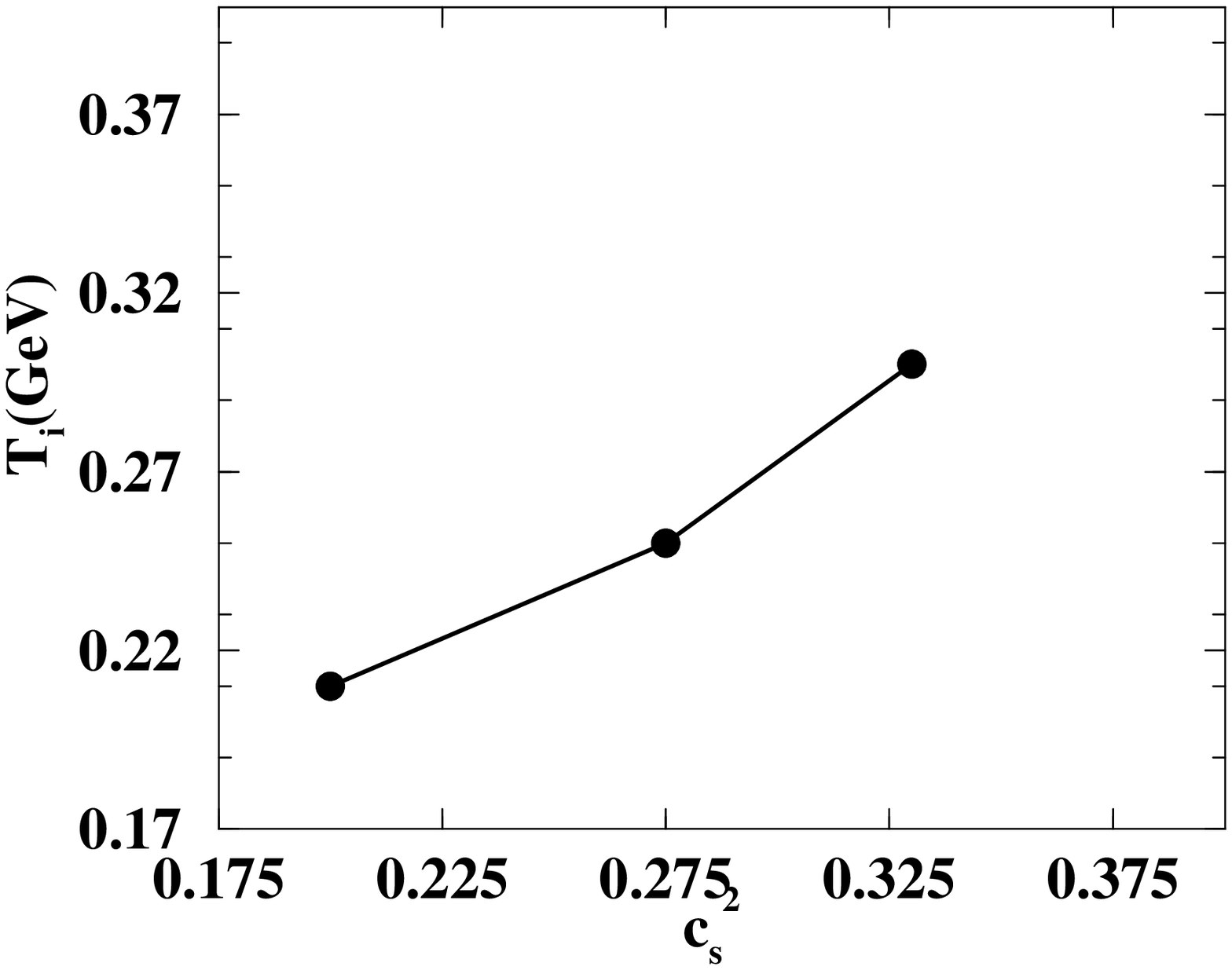}
\caption{The variation of $T_i$ with $c_s^2$ for fixed $dN/dy$.}
\label{FIG7}
\end{center}
\end{figure}

In Fig.~\ref{FIG7} we show the variation of $T_i$ with $c_s^2$
obtained by constraints imposed by the 
experimental data on $R_{\mathrm AA}$ and $dN/dy$.
The value of $T_i$ varies from 210 to 300 MeV depending on the
value of $c_s$. It is interesting to note that the lowest value of $T_i$ obtained from the present
analysis is well above the quark-hadron phase transition temperature, 
indicating the fact that the system  formed in Au+Au collisions
at $\sqrt{s_{\mathrm NN}}=200$ GeV might be formed 
in the partonic phase.

\begin{figure}[h]
\begin{center}
\includegraphics[scale=0.43]{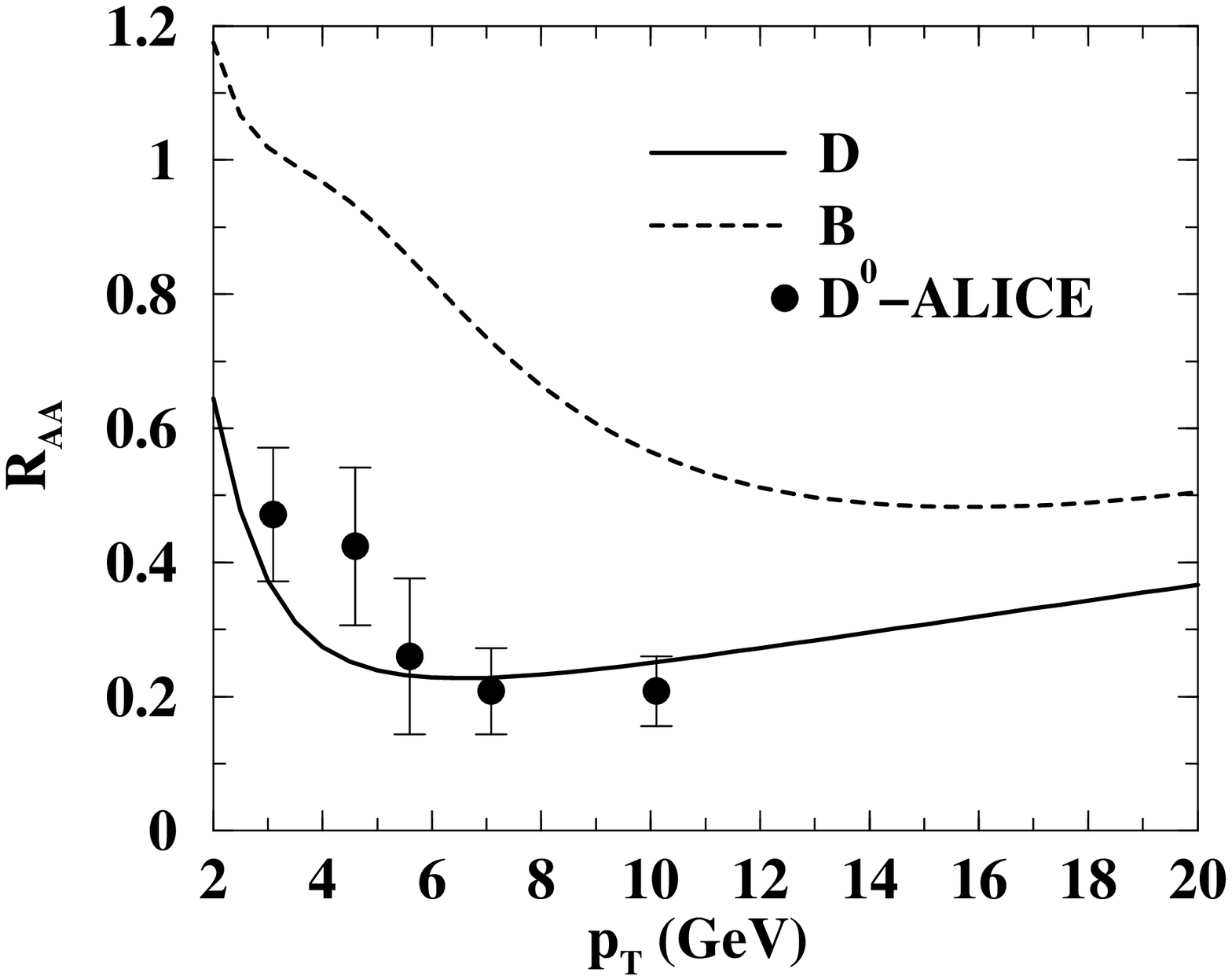}
\caption{$R_{AA}$ as a function of  $p_T$ for $D$ and $B$ mesons  at LHC.
Experimental data taken from ~\cite{ALICE}.}
\label{FIG8}
\end{center}
\end{figure}
The data for $R_{AA}$ of $D^0$ from ALICE are also contrasted with our results (see fig.\ref{FIG8})
when $c_s^2=1/3$
[\ref{FIG8}]. The EoS sets the expansion time scale for the system as $\tau_{\mathrm exp}\sim [(1/\epsilon)d\epsilon/d\tau]
^{-1}\sim\tau/(1+c_s^2)$  indicating the fact that lower value of $c_s$ 
makes the expansion time scale longer {\it i.e.} the rate of expansion slower.
This issue is further discussed in details in \cite{mazumderpaper2}.

The result of $R_{AA}$, where the EoS contains temperature dependent $c_s^2$, is
displayed in Fig.~\ref{FIG4}. Here, the experimentally measured suppression \cite{phenixe,stare}
has been reproduced reasonably well at $T_i=250$ MeV and $\tau_i= 0.83$ fm/c. The $T_i$
value, however, may increase if we take into account
the transverse expansion because the inflation dilutes the 
medium. Now, we know that larger $c_s$ makes the QGP life time smaller leading to lesser
suppression of HQ propagating through QGP for a shorter time. Therefore, when the value of 
$c_s^2$ is taken to be 1/3, {\it i.e.} the highest possible value, the maximum value of
$T_i$, which is in this analysis 300 MeV, will be reached. The value of $s_i$ is$\sim 59/$
fm$^{3}$ at $T_i$. These values of $T_i$ and $s_i$ are considered as the highest
values of them admitted by the data. The result for the highest value of $c_s^2$ is illustrated in Fig.~\ref{FIG5}.
Likewise, the result of $R_{AA}$ in the case when $c_s^2=1/5$ is depicted in Fig.~\ref{FIG6}. In 
this case, the data is well reproduced at $T_i=210$ MeV and $s_i\sim 19.66/$fm$^3$.

Now, it is expected that the central rapidity
region of the system formed after nuclear collisions
at high energy RHIC and LHC run is almost net baryon free. Therefore,
the equation governing the conservation of net baryon number need not be considered here
and all our calculations are valid for zero baryonic chemical potential ($\mu_B=0$)
cases. One may be interested in calculating transport coefficients in $\mu_B\neq0$
case which may be of importance in low energy RHIC run \cite{lowephenix,lowestar} and GSI-FAIR
\cite{cbm}. This aspect is discussed in the next section.

\section{Drag and diffusion at finite baryonic chemical potential }
The nuclear collisions at low energy RHIC run~\cite{lowephenix,lowestar}
and GSI-FAIR~\cite{cbm}  is expected to
create a thermal medium with  large baryonic chemical potential ($\mu_B$)
and moderate temperature ($T$). So the effect of baryonic chemical potential ($\mu_B$) 
on the transport coefficients of HQ should also been taken into account.
Both the temperature ($T$) and quark chemical potential, $\mu$ ($=\mu_B/3$) dependence of 
drag  enter through the thermal distribution.

\begin{figure}
\begin{center}
\includegraphics[scale=0.43]{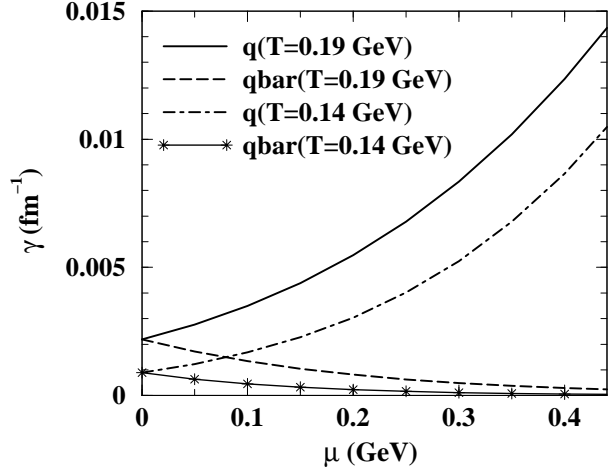}
\caption{Variation of the drag coefficient of charm quark due to its
interactions with light quarks and anti-quarks as a 
function of  $\mu$ for different temperatures.}
\label{Fig1}
\end{center}
\end{figure}

\begin{figure}
\begin{center}
\includegraphics[scale=0.43]{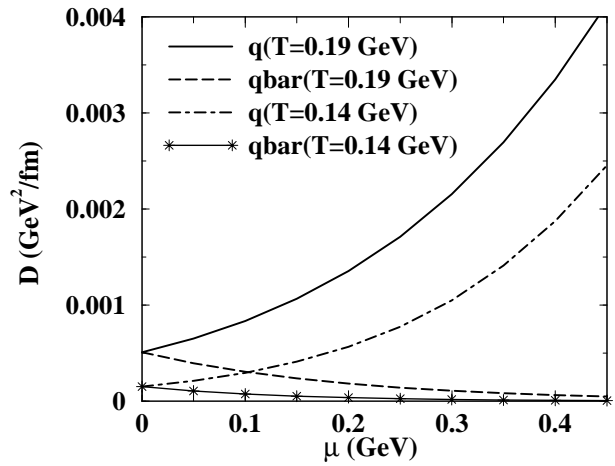}
\caption{Variation of the diffusion coefficient of charm quark due to its
interactions with light quarks and anti-quarks as a 
function of  $\mu$ for different temperatures.}
\label{Fig2}
\end{center}
\end{figure}

The variation of the drag coefficients of charm quarks 
(because of its interactions with quarks and anti-quarks) 
with the baryonic chemical 
potential  for different $T$ are displayed in Fig.~\ref{Fig1}.
The drag coefficient for the process : $Q$g$\rightarrow Q$g
is  $\sim 8.42\times 10^{-3}$ fm$^{-1}$ 
($1.86\times 10^{-2}$ fm$^{-1}$) for  $T=140$ MeV (190 MeV)
(not shown in Fig.~\ref{Fig1}).
The dependence of the drag and diffusion coefficients on temperature ($T$) and baryonic 
chemical potential ($\mu$) may be understood as follows:
As already discussed, the drag may be defined as the thermal average of the
momentum transfer, $p-p'$, weighted by the square of the invariant transition amplitude 
for the reactions $qQ\,\rightarrow\, qQ$,  
$Q\bar{q}\,\rightarrow\, Q\bar{q}$ and
$gQ\,\rightarrow\,gQ$. 
The average momentum of the quarks of the thermal bath increases with both T and $\mu$. 
The increase in average momenta enables the thermal quarks to transfer larger momentum; 
and thus enhances the drag coefficient.
This trend is clearly observed 
in the results displayed in Fig.~\ref{Fig1} for 
charm quark.  The drag due to the process $Qq\rightarrow Qq$ is 
larger than the  $Q\bar{q}\rightarrow Q\bar{q}$ interaction
because for $\mu\neq0$.  The $Q$ propagating
through the medium interacts with more $q$ than $\bar{q}$ at a given $\mu$
;and so, zero chemical potential case yields the same contributions from quarks and 
anti-quarks.  

\begin{figure}
\begin{center}
\includegraphics[scale=0.43]{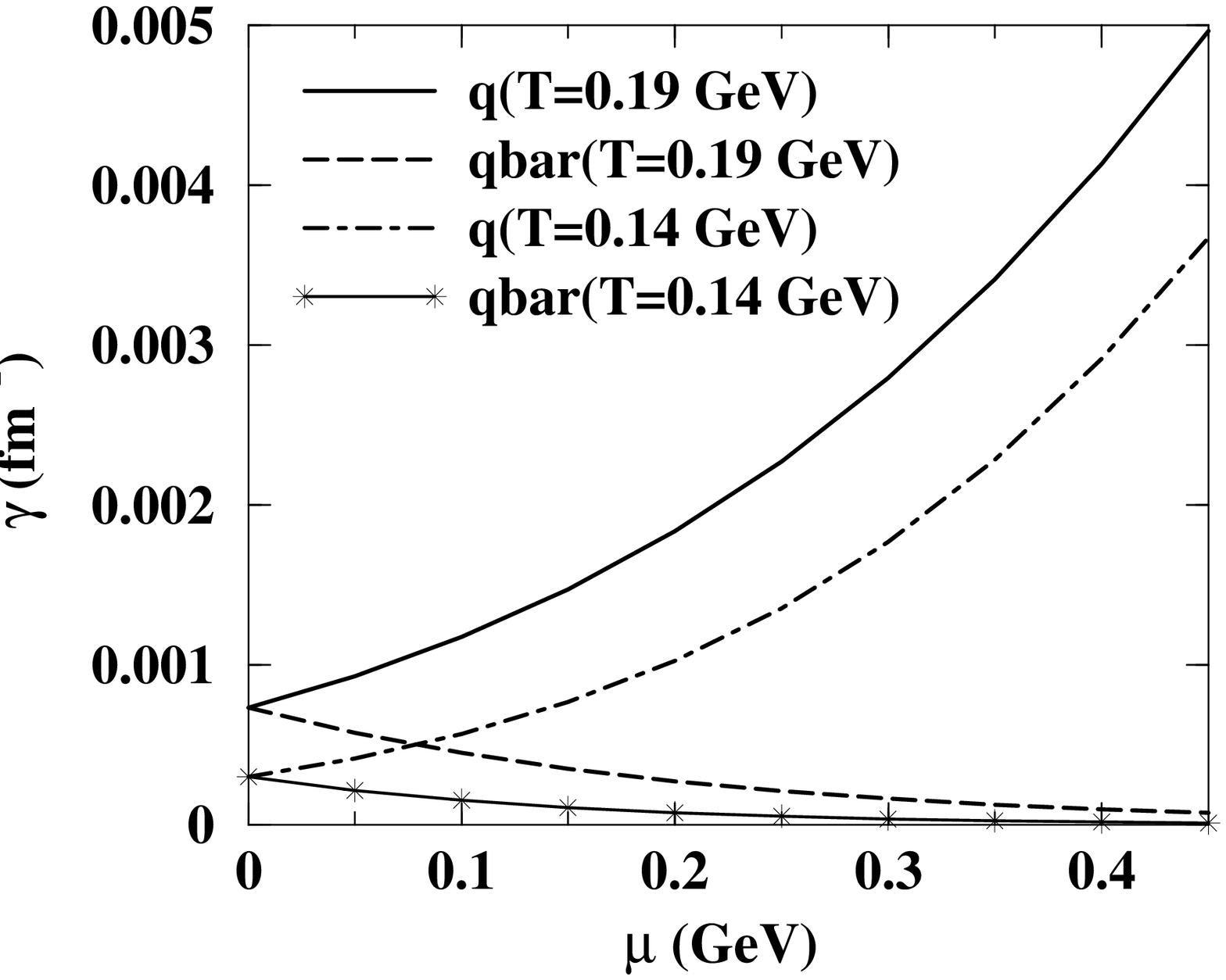}
\caption{Same as Fig.~\protect{\ref{Fig1}} for bottom quark}
\label{Fig3}
\end{center}
\end{figure}

\begin{figure}
\begin{center}
\includegraphics[scale=0.43]{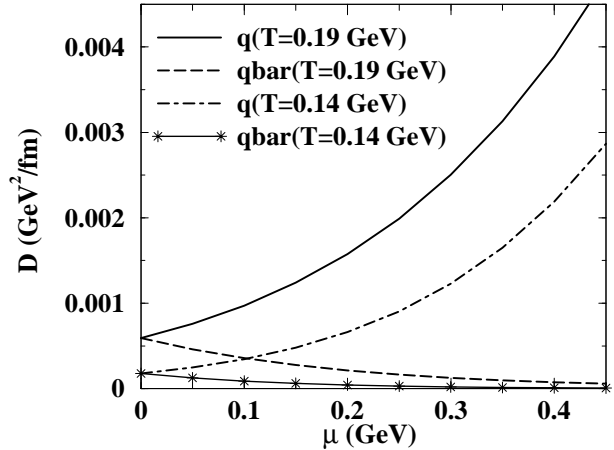}
\caption{Same as Fig.~\protect{\ref{Fig2}} for bottom quark.}
\label{Fig4}
\end{center}
\end{figure}

Similarly, it may
be argued that the diffusion coefficient, which involves the square of
the momentum transfer, should also increase with
$T$ and $\mu$ as observed in Fig.~\ref{Fig2}. The diffusion
coefficient for charm quarks due to its interaction
with gluons is given by
 $\sim 1.42\times 10^{-3}$ GeV$^2$/fm 
($4.31\times 10^{-3}$ GeV$^2$/fm) for $T=140$ MeV (190 MeV).
The drag and diffusion coefficients for bottom quarks are
displayed in Figs.~\ref{Fig3} and ~\ref{Fig4} respectively,
showing qualitatively similar behaviour with that of charm quarks.
The drag coefficients for bottom quarks due to the process 
$Q$g$\rightarrow Q$g
is given by $\sim 3.15\times 10^{-3}$ fm$^{-1}$ and   
$6.93\times 10^{-3}$ fm$^{-1}$ at $T$ = 140 MeV and 190 MeV respectively.
The corresponding diffusion coefficients 
are $\sim 1.79\times 10^{-3}$ GeV$^2$/fm  and
$5.38\times 10^{-3}$ GeV$^2$/fm at T=140 MeV and 190 MeV respectively. 
The chemical potential dependent drag and diffusion coefficients will be used later 
to find out the nuclear suppression of the heavy flavours for 
low energy RHIC experiments. 

\section{Nuclear Suppression in Baryon Rich QGP}
For low energy collisions, the collisional energy loss of heavy
quarks dominate over the radiation because HQs are produced with very low momentum. 
Moreover, the thermal production of charm and
bottom quarks can be ignored in the range of temperature and
baryonic chemical potential  considered in the low energy nuclear collisions. 
Therefore, one can apply the FP equation for non-zero $\mu_B$
for the description of HQ evolution in the baryon rich QGP. 
The drag and diffusion
coefficients are, now, functions of both the thermodynamical variables:
$\mu_B$ and $T$.

\begin{figure}
\begin{center}
\includegraphics[scale=0.43]{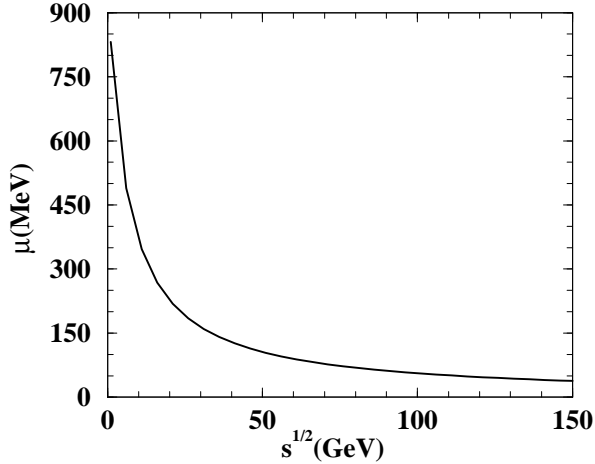}
\caption{Variation of $\mu_B$ with respect to the $\sqrt{s_{NN}}$. }
\label{chep}
\end{center}
\end{figure}

The colliding energy ($\sqrt{s_{NN}}$) dependence of the chemical potential 
can be obtained from the parametrization of the  experimental data on
hadronic ratios as~\cite{ristea} (see also~\cite{andronic}),
\be
\mu_B(s_{NN})=a(1+\sqrt{s_{NN}}/b)^{-1}
\label{mub1}
\ee
where $a = 0.967\pm 0.032$ GeV and $b=6.138 \pm 0.399$ GeV.
The parametrization in Eq.~\ref{mub1} gives the values
of $\mu_B$.
The chemical potential of the system, at mid rapidity, decreases with 
respect to the colliding energy as shown in Fig.~\ref{chep}. So the 
composition of matter produced at LHC and RHIC is different from the
matter produced at low energy RHIC run. At LHC and RHIC the matter 
produced at the mid rapidity is almost net baryon free, whereas the matter 
produced at the low energy RHIC run are dominated by baryons at the mid rapidity.

To take care 
of this extra baryons we need to solve the energy-momentum conservation equation
constrained by the baryon-number conservation equation. 
i.e. we simultaneously solve 
\be
\partial_\mu T^{\mu\nu}=0 
\ee
and
\be
\partial_\mu n_B^\mu=0
\ee
in (1+1) dimension with boost invariance along the longitudinal
direction~\cite{jdbjorken}. 
In the baryon-number conservation equation 
$n_B^\mu=n_Bu^\mu$ is the baryonic flux 
and $u^\mu$ is the hydrodynamic
four velocity. The initial baryonic chemical potential 
carried by the  quarks $\mu (=\mu_B/3)$ are shown in Table~\ref{ttt} for 
various $\sqrt{s_{NN}}$ under consideration. 

\begin{table}[ht]
\centering
\caption{The values center of mass energy $\sqrt{s_{NN}}$ , $dN/dy$, initial temperature ($T_i$)
and quark chemical potential  - used
in the present calculations.}
\begin{tabular}{lccr}
\hline
$\sqrt(s_{\mathrm NN})$(GeV)&$\frac{dN}{dy}$ &$T_i$(MeV) &$\mu$(MeV)\\
\hline
39&617 &240  &62\\
27&592 &199  &70\\
17.3&574 &198  &100\\
7.7 &561 &197  &165\\
\hline
\label{ttt}
\end{tabular} 
\end{table}

Multiplicities for various $\sqrt{s_{NN}}$
have been calculated from the Eq. below~\cite{kharzeevnardi};
\be
\frac{dN}{dy}=\frac{dn_{pp}}{dy}\left[(1-x)\frac{<N_{part}>}{2}+x<N_{coll}>\right]
\ee
$N_{coll}$ is the number of
collisions and contribute  $x$ fraction to the multiplicity $dn_{pp}/dy$
measured in $pp$ collision. The number of participants,
$N_{part}$ contributes a 
fraction $(1-x)$ to $dn_{pp}/dy$, which is given by 
\be
\frac{dn_{pp}}{dy}=2.5-0.25ln(s)+0.023ln^2(s)
\ee
The values of the number of participants, $N_{part}$,
and number of collisions, $N_{coll}$,  are estimated for $(0-5\%)$ centrality's by using Glauber Model~\cite{Gla}.
The value of $x$ depends very weakly on $\sqrt{s_{NN}}$~\cite{bbback},
in the present work we have taken $x=0.1$ for all the colliding energies considered 
herein.

\begin{figure}[ht]
\begin{center}
\includegraphics[scale=0.43]{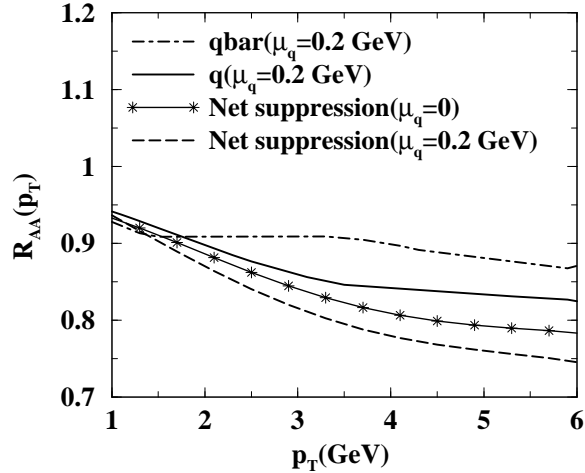}
\caption{The nuclear suppression factor $R_{AA}$ as a function
of $p_T$ due to the interaction of the charm quark (solid line)
and anti-quark (dashed-dot line) for $\mu=200$ MeV.
The net suppressions including the interaction of quarks, anti-quarks and
gluons for $\mu=200$ MeV (dashed line) and $\mu=0$ (with asterisk) 
are also shown.}
\label{fig20}
\end{center}
\end{figure}

\begin{figure}[ht]
\begin{center}
\includegraphics[scale=0.43]{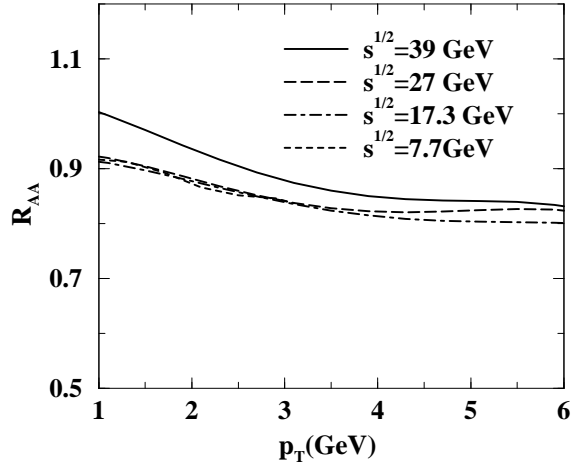}
\caption{Nuclear suppression factor, $R_{\mathrm AA}$ as function of $p_T$
for various $\sqrt{s_{NN}}$.}
\label{fig21}
\end{center}
\end{figure}

\begin{figure}[ht]
\begin{center}
\includegraphics[scale=0.43]{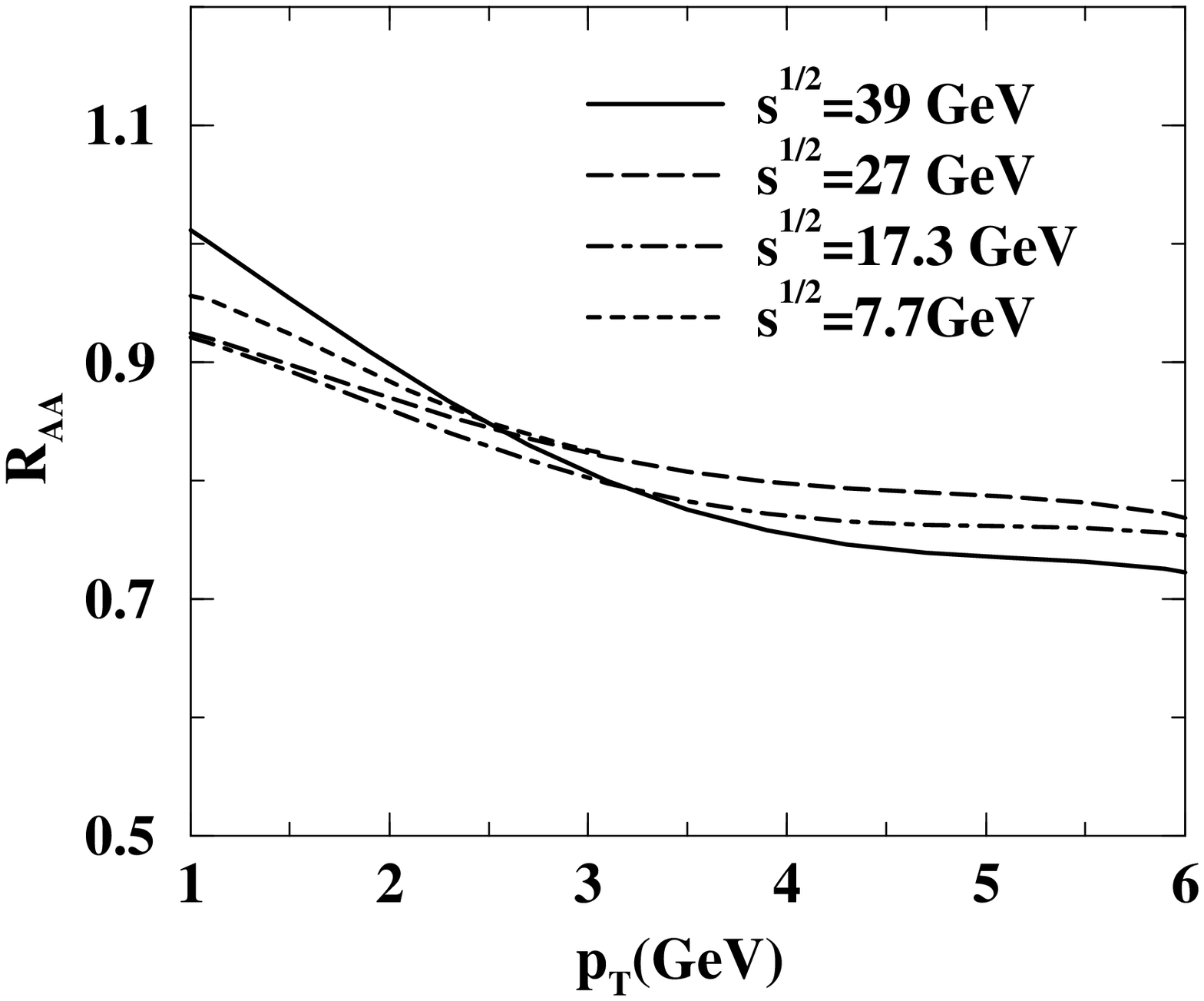}
\caption{Same as Fig.~\protect{\ref{fig21}} with enhancement of cross section 
by a factor of 2.}
\label{fig22}
\end{center}
\end{figure}

We need the initial heavy quark momentum distributions 
for solving the FP equation at finite baryonic chemical potential.  
For low collision energy rigorous QCD based calculations for 
heavy flavour production at proton-proton collision is not available. In the present work 
the initial HQ distribution  is obtained from pQCD calculation~\cite{combridge,pqcd}
for the processes: $gg\rightarrow Q\bar{Q}$
and $q\bar{q}\rightarrow Q\bar{Q}$.
To highlight the effect of non-zero baryonic chemical
potential we evaluate $R_{\mathrm AA}$ for $\mu=200$ MeV
and $\mu=0$ for a given $T_i=200$ MeV. The
results are shown in Fig.~\ref{fig20}
representing the combined effects of 
temperature and baryon density on the drag and diffusion coefficients. 
The drag coefficient of the heavy quarks due to its interaction with
quarks is larger than that owing to its interactions with the anti-quarks
(Fig.\ref{Fig1}), resulting in larger suppression in the former case than the latter.
The net suppression of the electron spectra from the Au+Au collisions 
compared to p+p collisions is affected by quarks, anti-quarks and gluons.
The results for net suppressions are displayed for $\mu=200$ MeV (dashed
line) and $\mu=0$ (with asterisk). 
The experimental detection of the
non-zero baryonic effects will shed light on the net baryon density
(and hence baryon stopping).
However, whether the effects of non-zero baryonic 
chemical potential is detectable or not will depend on the 
overall experimental performance.

The results for $R_{AA}$ are shown in Fig.~\ref{fig21}
for various $\sqrt{s_{NN}}$ with inputs from Table~\ref{ttt}.
We find that at large $p_T$ the suppression is similar for all energies
colliding energies considered in the present work. This is because the collisions
at high $\sqrt{s_{NN}}$  are associated with large temperature but
small baryon density at mid-rapidity- which is compensated by  large 
baryon density and small  temperature at low $\sqrt{s_{NN}}$ collisions.  
Low $p_T$ particles predominantly originate from low temperature and 
low density part of the evolution where drag is less and so is the nuclear suppression.

In our earlier work~\cite{das} we have evaluated the 
$R_{\mathrm AA}$ for non-photonic single electron spectra 
resulting from the semileptonic decays of hadrons containing 
heavy flavours. We observed that the data from RHIC collisions
at $\sqrt{s_{NN}}=200$ GeV are well reproduced by enhancing the
pQCD cross sections by a factor 2  and with an equation of state
$P=\epsilon/4$ for collisional loss. In the same spirit we evaluate $R_{\mathrm AA}$ 
with doubled pQCD cross section
and  keeping all other quantities same~(Fig.~\ref{fig22}).
The results in Fig.~\ref{fig22} show stronger  suppression
as compared to the results displayed in Fig.~\ref{fig21}, however
it is similar for all the colliding energies considered in the present article. 

\section{Glimpses of Elliptic Flow }
However, apart from the nuclear suppression factor $R_{AA}$, another 
experimental observable of heavy flavour, elliptic flow ($v_2$), can be studied 
within the framework of Fokker-Planck equation.
The elliptic flow ($v_2$) of the
produced particles has been considered as one of 
the most promising signals for the early thermalization 
of the matter formed in heavy ion collision. 
If we consider a thermalized ellipsoidal spatial domain
of QGP, originated due to non-central nucleus nucleus collisions, 
with major and minor axes of lengths $l_y$ and  $l_x$ 
(determined by the collision geometry) respectively then the 
pressure gradient is larger along the minor axis compared to 
that along the major axis because $l_y > l_x$. Now, pressure gradient
is force and hence force along the minor axis is larger than that along the
minor axis. Consequently, the HQ moves faster in this direction. 
Therefore, the momentum distribution of electrons
originating from the decays of charmed hadrons ($D$ mesons) produced
from the charm quark fragmentation will be anisotropic; and since 
$v_2$ is the second Fourier's coefficient of the momentum
distribution, the spatial anisotropy
is thus reflected in the momentum space anisotropy.

$v_2$ is  
sensitive to the initial conditions and the equation of state 
(EoS) of the evolving matter formed in heavy ion
collision~\cite{hiranolec,Huovinen_v2,Teaney_v2}. 
Several theoretical attempts has been made in this direction 
to study both the nuclear suppression factor and elliptic flow 
of the heavy flavour within a single framework. Although several 
authors have reproduced the large suppression 
of high momentum heavy quarks  reasonably well yet it is still difficult to explain 
HQ elliptic flow simultaneously within the same 
set of initial parameters. It is also imperative
to mention that no calculation with radiative energy loss is
still able to reproduce heavy flavour elliptic flow. After this brief discussion,
we refer the interested readers to
Refs.~\cite{moore,alberico,bass,hees,gossiaux,rappv2,hvh2,hirano,skd} for
more interesting aspects of elliptic flow.

\section{Summary and Discussions:}
In summary, we have tried to give a systematic account
of evaluating collisional and radiative transport coefficients of heavy quarks
passing through quark gluon plasma. 
The nuclear suppression factor ($R_{AA}$)
of heavy quarks has been evaluated using Fokker-Planck equation.  
The present article confines the discussions within perturbative QCD.
Recently, ref. \cite{adscftskdad} finds out the drag force and $R_{AA}$
of charm quarks propagating through
thermalized QGP within the framework of both AdS/CFT and AdS/non-CFT. 
Interested readers are referred to \cite{gubser,giecold} for similar discussions.

We have studied the momentum dependence of transport coefficients
and observe that the momentum dependence is crucial in reproducing the trend in
the $p_T$ dependence of the experimental data. Also,
the dead-cone factor ($F_{DC}$) weighted by the  
Gunion-Bertsch spectrum for radiated gluon from heavy quarks is implemented. 
It is worth mentioning that though the present article speaks at length about elastic as well
as bremsstrahlung energy loss of HQs and tells about dominance 
of the latter in high $p_T$ region; yet there are a plethora of articles
which show the importance of elastic energy
loss in context of experimental observables in entire momentum range attainable
by HQs (\cite{moore,rappv2,mustafaprcsingle}). This issue, as a matter of fact,
still awaits an unambiguous settlement.

Momentum independent transport coefficients in non-zero baryonic
chemical potential region has also been studied and theoretical
estimates of the suppression factor is provided. In the Appendix,
we have discussed the calculation of elastic transport coefficients
within the ambit of QCD hard thermal loop perturbation theory. Also,
radiative suppression due to off-shellness of produced partons, which
we may call the `dead cone due to virtuality', has also been discussed.
This effect of off-shellness of heavy quarks as well as 
HTL calculations may be extended to encompass radiative processes 
which, as we have already discussed, will play the most significant 
part in high energy regime. The calculations of $R_{AA}$
including the aspects just mentioned may be contrasted with
the experimental data.


Though \cite{moore,das,san,dams,alberico,dca,gossiaux,rappv2,hirano} 
attempt to study $R_{\mathrm AA}$ and $v_2$ of the heavy quark, yet the role of 
hadronic matter has been ignored. However, to make the
characterization of QGP reliable, the role of the
hadronic phase should be taken into consideration 
and its contribution must be subtracted out from the observables.  
Although a large amount of work has been done 
on the diffusion of heavy quarks in QGP, the diffusion of
heavy mesons in hadronic matter has received much
less attention so far. Recently the diffusion coefficient 
of $D$ meson has been  evaluated  using heavy meson 
chiral perturbation theory~\cite{laine} and also by using the empirical 
elastic scattering amplitudes~\cite{MinHe} of $D$ mesons with thermal hadrons.
The interactions of D meson with pions, nucleons,
kaons and eta particles have been evaluated using Born amplitudes~\cite{Ghosh} 
and unitarized chiral effective $D\pi$ interactions~\cite{abreu}. 
D and B-meson scattering lengths have also been used as dynamical input 
to study the drag and diffusion coefficients~\cite{Sarkar,cabrera}. All these studies observed
that the magnitude of both the transport coefficients are significant, indicating substantial amount
of interaction of the heavy mesons with the thermal bath. The results may have significant impact
on the experimental observables like the suppression of single electron spectra \cite{dgsa} originating from the
decays of heavy mesons produced in nuclear collisions at RHIC and LHC energies.


\vskip 0.6in

~~~~~~~~~~~~~~~~~~~~~~~~~~~~~~~~~~~~~~~~~~~~~~ {\large \bf Appendix}

{\section*{Hard Thermal Loop (HTL) approximations and transport coefficients:}}
We have used $T=0$ pQCD matrix elements for calculating
collisional transport coefficients so far. The $t$ channel divergence
due to soft intermediary gluon exchange has been shielded by an ad hoc 
replacement of $t$ by $t-m_D^2$, where $m_D$ is the (static) Debye
screening mass of gluon. Here we use thermal
resummed (HTL) gluon propagators \cite{Bellac,kap,heisel} for calculating the matrix elements
for the processes like $Qq\rightarrow Qq$ and $Q$g$\rightarrow Q$g, where
$Q(q)$ stands for heavy(light) quark and $g$ stands for gluon, for a self-consistent
shielding of $t$ channel divergence. Since the light quarks and gluons of thermal 
bath are hard, {\it i.e.} their momenta are $\sim T$, we may neglect the vertex correction
(assuming $g<<1$) arising due to ggg or $qq$g vertex. The effect of full gluon spectral
function on the collisional drag coefficient will be investigated for
HQs. 

\subsection*{HTL gluon propagator:}
\begin{figure}[h]      
\begin{center}                                                               
\includegraphics[scale=0.9]{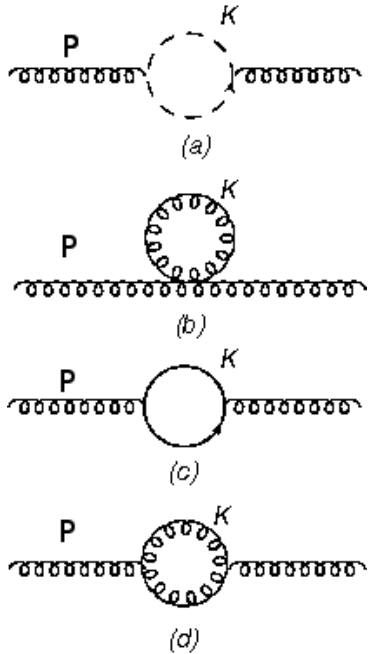}                                 
\end{center}                                                                 
\caption{Feynman diagrams contributing to gluon self-energy up to one loop. (a)ghost-gluon loop. 
(b)four-gluon vertex. (c) quark-antiquark pair creation. (d)three-gluon vertex.}
\label{figselfgluon}                                                               
\end{figure}

For finding out the HTL gluon propagator ($\Delta^{\mu\nu}$) we need
HTL approximated self-energy of gluon which goes as an input to  
$\Delta^{\mu\nu}$ to be used as effective thermal propagator regularizing 
the $t$ channel divergence. The gluon self-energy in HTL approximation is 
discussed in detail in Ref.~\cite{Bellac,kap}. 
In this section we  give only an outline of the scheme. 
There are four diagrams which contribute to gluon  self-energy~(Fig.~\ref{figselfgluon}).
The loop integrations can be written down easily if we keep in mind that the loop-momentum, $K=(k_0,\vec{k})$
is much larger compared to external gluon momentum, {\it i.e.} $K>>P$ which
enables us to use simplified ggg vertex \cite{Bellac}. Our goal will be to find out $T^2$
contributions of self-energy which is the leading behaviour of self-energy in terms
of temperature, $T$. This $T^2$ contribution of self-energy is called the `hard thermal loop'
contribution. We may argue that the hard thermal loop contribution of the gluon self
energy is due to momenta $\sim T$ if we look at the generic integral of the type

\be
\int_0^{\infty} k f(k) dk=\frac{\pi^2 T^2}{12},
\label{avk}
\ee
which appears during self-energy calculation. $f$ in Eq. \ref{avk} is the distribution function and
the leading contribution in Eq. \ref{avk} is given by $k\sim T$. Henceforth, we obtain
a scale of momentum $\sim T$, which will be called `hard' compared to another `soft'
scale $\sim gT$, where $g(<<1)$ is the colour charge. HTL gluon propagators
are to be used for processes exchanging very soft gluons. As the $t$ channel divergence 
is equivalent to dominance of soft gluon exchange processes, the use of HTL propagator
is justified and consistent.


Now, let us we define the following useful quantities~\cite{Bellac}
required to write down the gluon propagator 
in thermal medium. Let $u_{\mu}$ be
the fluid four-velocity, with normalization condition $u^\mu u_\mu=1$. The
fluid four-velocity gives rise to two directions; and so any four-vector 
$P^\mu$ can be decomposed into components parallel and perpendicular to the fluid velocity:
\bea
\omega=P.u\nn\\
\tilde{P}_{\mu}=P_{\mu}-u_{\mu}(P.u)\nn\\
\label{energy}
\eea
where
\bea
P^2=\omega^2-p^2 \nn\\
\tilde{P}^2=-p^2
\label{fourmomsq}
\eea
Eqs. \ref{energy} and \ref{fourmomsq} are valid in the local rest frame of fluid, i.e. in a frame where
$u=(1,\vec{0})$. Similarly a tensor orthogonal to $u_{\mu}$ can be defined as,
\be
\tilde{g}_{\mu \nu}=g_{\mu \nu}-u_{\mu}u_{\nu}
\ee
The longitudinal and transverse projection tensors, $\mathcal{P}_L^{\mu\nu}$ 
and  $\mathcal{P}_T^{\mu\nu}$ respectively, are defined as~\cite{heisel}
\bea
\mathcal{P}^{\mu\nu}_L=-\frac{1}{P^2 p^2} (\omega P^{\mu}-P^2 u^{\mu})(\omega P^{\nu}-P^2 u^{\nu})
\label{pl}
\eea
\bea
\mathcal{P}^{\mu\nu}_T=\tilde{g}_{\mu \nu}+\frac{\tilde{P}_{\mu}\tilde{P}_{\nu}}{p^2}
\label{pt}
\eea
which are orthogonal to $P^{\mu}$ as well as to each other, {\it  i.e.}
\be
P_{\mu}\mathcal{P}^{\mu\nu}_L=P_{\mu}\mathcal{P}^{\mu\nu}_T=\mathcal{P}^{\mu}_{L\nu}\mathcal{P}^{\nu\rho}_T=0
\label{ortho}
\ee
But,
\be
\mathcal{P}^{\mu\rho}_i\mathcal{P}_{i\nu\rho}=\mathcal{P}_{i\nu}^{\mu}~~~,i=L, T
\label{contract}
\ee
The HTL gluon propagator, which is given by: 
\bea
\Delta^{\mu\nu}=\frac{\mathcal{P}_T^{\mu\nu}}{-P^2+\Pi_T}+\frac{\mathcal{P}_L^{\mu\nu}}{-P^2+\Pi_L}
+(\alpha-1)\frac{P^{\mu}P^{\nu}}{P^2}
\label{htlprop}
\eea
will need HTL approximated longitudinal and transverse self-energies $\Pi_L$ and $\Pi_T$
respectively, too.  $\Pi_L$ and $\Pi_T$ are given by
\bea
\Pi_L(P)=(1-x^2)\pi_L(x),~~~~\Pi_T(P)=\pi_T(x)
\label{cappil}
\eea
where $x=\omega/|\vec{p}|$ ($P\equiv(\omega,\vec{p})$, see Fig.~\ref{figselfgluon})
and scaled self-energies $\pi_T$ and $\pi_L$ are given by~\cite{Bellac}, 
\bea
\pi_T(x)=
m_D^2\left[\frac{x^2}{2}+\frac{x}{4}(1-x^2)ln \left(\frac{1+x}{1-x}\right) 
-i\frac{\pi}{4}x(1-x^2)\right]\nn\\
\eea
\bea
\pi_L(x)=m_D^2\left[1-\frac{x}{2}ln (\frac{1+x}{1-x})+i \frac{\pi}{2}x\right]\nn\\
\label{pil}
\eea
where $m_D$ is the thermal mass of gluon; and
is given by $m_D^2=g^2T^2(C_A+N_f/2)/6$, where $C_A=3$ is the Casimir 
of adjoint representation of SU(3) and $N_f=2$ is the number of flavours.

\subsection*{Finding out matrix elements for Qq scattering:}
\begin{figure}[h]      
\includegraphics[scale=0.6]{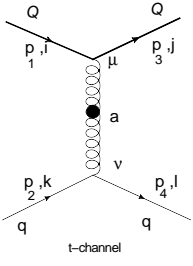}                                 
\caption{$Qq\rightarrow Qq$ Feynman diagram. Bold lines are for heavy quarks(Q).}
\label{Qq}                                                               
\end{figure}                                                                 

From Fig. \ref{Qq} we can calculate the t-channel matrix element for the process $Qq\rightarrow Qq$.
We will use the HTL gluon propagator \cite{Bellac}. Pictorially, an HTL
propagator will be denoted by a solid circle. We can write the amplitude 
in Feynman Gauge ($\alpha=1$) from Fig. \ref{Qq} as,

\bea
-iM_t= \overline{u}(p_3)(-ig\gamma^{\mu}t^a_{ji})u(p_1)\left[-i\Delta_{\mu\nu}\right]
\overline{u}(p_4)(-ig\gamma^{\nu}t^a_{lk})u(p_2)
\label{Qqtch}
\eea
where $g$ is strong coupling and $g^2=4\pi\alpha_s$. $i,j,k,l~~(i\neq j,~k\neq l)$ are quark colours and 
`a' is the colour of intermediary  gluon with polarizations $\mu,\nu$. After squaring and averaging
over spin and colour as well as using eq. \ref{htlprop} we get, 

\bea
\frac{\left|M_{Qq}\right|^2}{4 C_{Qq} g^4}&=&2\frac{p_4.\mathcal{P}_T.p_3 p_2.\mathcal{P}_T.p_1}{\left(t-\Pi_T\right)^2}
+2\frac{p_4.\mathcal{P}_L.p_3 p_2.\mathcal{P}_L.p_1}{\left(t-\Pi_L\right)^2}
+2\frac{p_4.\mathcal{P}_T.p_1 p_2.\mathcal{P}_T.p_3}{\left(t-\Pi_T\right)^2}\nn\\
&+&2\frac{p_4.\mathcal{P}_L.p_1 p_2.\mathcal{P}_L.p_3}{\left(t-\Pi_L\right)^2}
+2A \frac{p_4.\mathcal{P}_L.p_3 p_2.\mathcal{P}_T.p_1+p_4.\mathcal{P}_T.p_3 p_2.\mathcal{P}_L.p_1}
{\left(t-\Pi_T\right)^2 \left(t-\Pi_L\right)^2}\nn\\
&+&2A\frac{p_4.\mathcal{P}_L.p_1 p_2.\mathcal{P}_T.p_3+p_4.\mathcal{P}_T.p_1 p_2.\mathcal{P}_L.p_3}
{\left(t-\Pi_T\right)^2 \left(t-\Pi_L\right)^2}\nn\\
&-&2p_4.p_2\left[\frac{p_3.\mathcal{P}_T.p_1}{\left(t-\Pi_T\right)^2}
+\frac{p_3.\mathcal{P}_L.p_1}{\left(t-\Pi_L\right)^2}\right]\nn\\
&-&2p_3.p_1\left[\frac{p_4.\mathcal{P}_T.p_2}{\left(t-\Pi_T\right)^2}
+\frac{p_4.\mathcal{P}_L.p_2}{\left(t-\Pi_L\right)^2}\right]\nn\\
&+&p_3.p_1p_4.p_2\left[\frac{2}{\left(t-\Pi_T\right)^2}+\frac{1}{\left(t-\Pi_L\right)^2}\right]\nn\\
&+&m^2\left[2\frac{p_4.\mathcal{P}_T.p_2}{\left(t-\Pi_T\right)^2}
+2\frac{p_4.\mathcal{P}_L.p_2}{\left(t-\Pi_L\right)^2}\right]\nn\\
&-&m^2\left[2\frac{p_4.p_2}{\left(t-\Pi_T\right)^2}+\frac{p_4.p_2}{\left(t-\Pi_L\right)^2}\right]
\label{Qqtchsq}
\eea
     
where $C_{Qq}=\frac{2}{9}$ is the Color factor, $Q^2\equiv t=(p_1-p_3)^2$, 
$A=t^2-t(Re\Pi_T+Re\Pi_L)+Re\Pi_T\Pi_L^*$ and we have used the following relations,

\bea
p_1.\mathcal{P}_L.p_2=p_3.\mathcal{P}_L.p_4=
p_4.\mathcal{P}_L.p_1=p_2.\mathcal{P}_L.p_3
\eea
as well as,
\bea
\Delta^{\mu\rho}\Delta^{*\nu}_\rho&=&\frac{\mathcal{P}_T^{\mu\nu}}{\left(t-\Pi_T\right)^2}
+\frac{\mathcal{P}_L^{\mu\nu}}{\left(t-\Pi_L\right)^2}\nn\\
\left|\Delta\right|^2&=&\Delta^{\mu\nu}\Delta^*_{\nu_\mu}
=\frac{2}{\left(t-\Pi_T\right)^2}+\frac{1}{\left(t-\Pi_L\right)^2}\nn\\
\label{proprels}
\eea
which can be proven using Eqs. \ref{pl}, \ref{pt}, \ref{htlprop} in the rest frame of fluid element. 

The $Q$g$\rightarrow Q$g scattering contains $s$, $t$ and $u$ channel
diagrams and they can easily be written just like $Qq$ case. There are
$|M_t|^2$, $|M_s|^2$ and $|M_u|^2$ as well as the interference
of $s$, $t$, $u$ channels as opposed to \cite{thomaprdrapid}
where only $s$ and $u$ channel interference term
exists (besides $|M_t|^2$) because gluon momenta are assumed to be much larger 
than that of HQ.
\begin{figure}[h]      
\begin{center}                                                               
\includegraphics[scale=0.5]{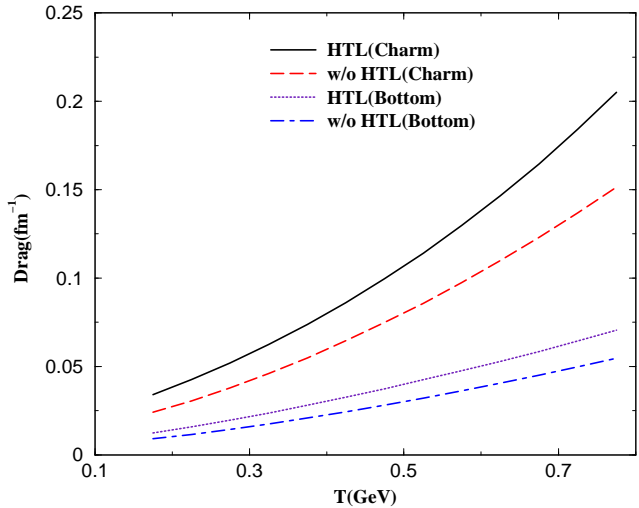}                                 
\end{center}                                                                 
\caption{ (Color online) Variation of drag of heavy quarks 
of momentum 1 GeV with temperature.}
\label{dragT}                                                               
\end{figure}                                                                 

\begin{figure}[h]      
\begin{center}                                                               
\includegraphics[scale=0.5]{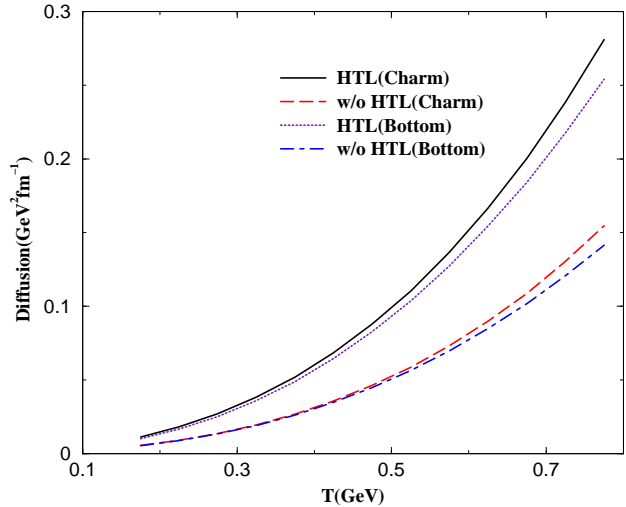}                                 
\end{center}                                                                 
\caption{ (Color online) Variation of diffusion of heavy quarks 
of momentum 1 GeV with temperature.}
\label{diffT}                                                               
\end{figure}                                                                 

\subsection*{Results and discussions on Collisional Drag and Diffusion coefficients using HTL propagator:}
We have found out the matrix elements for relativistic heavy quark
scattering elastically with light quarks and gluons of QGP
with arbitrary scattering angle. The variations of drags with temperature for HQs with momentum, $p=1$ GeV 
as a function of temperature are displayed in Fig.~\ref{dragT}. The results
clearly indicate an enhancement and rapid variation of drag using HTL propagator ($\gamma_{HTL}$)
compared to that in $T=0$ pQCD ($\gamma$).
The increase is more prominent for charm than bottom.   
We have explicitly checked that in the static limit 
the drag and diffusion using HTL propagator approaches that
in $T=0$ pQCD. $\gamma_{HTL}$ is greater than $\gamma$ for the entire momentum range considered here.
Again, drag being the measure of energy loss, increase in drag results in 
more suppression of heavy flavours measured at RHIC and LHC energies. 
From Fig. \ref{dragT} we observe that at 400 MeV temperature $\gamma_{HTL}$ 
for charm quark is $\sim$33\% more
than the $\gamma$. Whereas, the corresponding difference is $\sim$25\% for bottom quarks. We also observe
that this difference increases with the increase in temperature. 
Diffusion coefficients, plotted in fig. \ref{diffT},
 seem to be more sensitive to the use of effective propagator in a sense that we observe 
100\% change between the diffusion using HTL propagator ($D_{HTL}$) and that in $T=0$ pQCD ($D$)
at T = 400 MeV and this difference increases with $T$. 
Though unlike drag, this difference is not much (∼3.5\%) for a difference in charm and bottom quark masses.

\cite{thomaprdrapid} also calculates
the energy loss of heavy quarks using HTL propagator and
gets a drag which is 16\% less than that obtained in the present
paper at HQ momentum 4 GeV and $T=250$ MeV. 
\cite{moore} calculates diffusion coefficient of a non-relativistic
heavy quark in leading order as well as in next to leading order. The leading
order result surpasses the present result by 25\% at $T=300$ MeV and at a very low momentum
(0.2 GeV) of HQ.

However, radiative transport coefficients like drag are also needed for
we have seen that radiation becomes very important in high
energy regime. We can even extend the present calculation of collisional 
drag and diffusion using HTL propagator to radiative domains; but that
will increase the complexity of problem. 

\section*{Recent efforts of generalizing GB formula:}
Gunion-Bertsch formula is derived in the mid-rapidity region and 
there is another very recent development in this field proposed in ~\cite{uphoff} 
where a generalized Gunion-Bertsch formula for arbitrary forward and backward rapidity
region has been derived. This modification will be needed when one needs to compute
cross sections and rates. Now, this calculation involves two parts 
(a) keeping a factor $(1-x)^2$, where x is the fraction of light cone momentum carried by
the emitted gluon and (b) combination of calculations obtained from both $A^+=0$ and $A^-=0$ gauge
conditions for emitted gluon polarization. The final form of $M_{qq' \rightarrow qq'g}$
can be written as follows:
\bea
|M_{qq' \rightarrow qq'g}|^2=12 g^2 |M_{qq'\rightarrow qq'}|^2_{sa} (1-\bar{x})^2
\times \frac{q_{\perp}^2}{k_{\perp}^2(\vec{q_{\perp}}-\vec{k_{\perp}})^2},
\label{uphoffspectrum}
\eea
where
\be
\bar{x}=\frac{k_{\perp}e^{|\eta|}}{\sqrt{s}},
\ee
where $\eta$ is the rapidity of the emitted gluon.
With this, the exact differential cross-section ($\frac{d\sigma}{d\eta}$) for the process 
$ qq' \rightarrow qq'g $ is shown to be reproduced by using Eq. \ref{uphoffspectrum} 
for all rapidity ranges.

In all these calculations, we tacitly assume that the incoming jet is hard enough so that
eikonal approximation (straight path) for it is always valid. But there has been a recent
attempt of relaxing the eikonal approximation in ~\cite{abireikonal} and
15-20\% suppression in the differential cross-section of $2\rightarrow 3$ processes for moderately
hard jets because of the noneikonal effects has been found. 

\section*{Dead cone effect revisited and other aspects of energy loss}
~\cite{abirdcone}
computes the HQ-LQ $\rightarrow$ HQ-LQ-g scattering amplitude for 
soft gluon emission and finds out a general expression for radiative suppression factor for 
dead cone effect. In the limit $m<<\sqrt{s}$ and $\theta\rightarrow 0$, they reproduce Eq.\ref{khardead}.
In the backward rapidity region the gluon emission does not depend on the mass
and in $\theta\sim\pm\pi$ region there is no suppression, in contrast with what
we get from ~\cite{khar}(see \cite{kmpaul} also). \cite{abirhqelossplb} evaluates
the HQ energy loss employing the generalized dead-cone factor in \cite{abirdcone}. They
report similar energy loss for both massless and massive quark jets.

However, the high energy quarks and gluons produced from the hard collisions of 
the partons from the colliding nucleons 
are off-shell and their colour fields are stripped off, {\it i.e.} they have no field
to radiate. Therefore, the
partons' virtuality creates its own dead-cone which may be large depending on
the magnitude of the virtuality. The forbidden zone around the direction of motion of the 
partons due to its virtuality will here be called virtual dead cone. 
If the virtuality does not  disappear before the hadronization of the 
QGP then the dead cone suppression due to virtuality will play a decisive role 
in QGP diagnostics by jet quenching.  The conventional
dead cone (due to the mass of the quark) becomes important when the
virtuality of the quarks reduces to zero. 

For the demonstration of suppression of soft gluon radiation due to virtuality, 
we can take up the $e^+ e^− \rightarrow Q\bar{Q}$g process, where Q is heavy quark. The spectrum
of the soft gluons emitted by the virtual quarks can be shown to be~\cite{bavirtdead}
\bea
F=4\beta^2 \left(\frac{\frac{V^4}{k_0^2 E^2}+\frac{4V^2}{k_0 E}+4sin^2 \theta}
{\left(\frac{V^4}{k_0^2 E^2}+\frac{4V^2}{k_0 E}+4(1-\beta^2 cos^2 \theta)\right)^2}\right),
\label{deadvirt}
\eea
where external quarks are assumed to be on the verge of being on-shell
so that Dirac's equation can be applied. V is the ‘virtuality parameter’
defined by the equation, $V^2=q^2-m^2$ where $q ^2$ is four-momentum square
of external virtual particles; and $m$ is its mass, if any.
$q ^2 = m^2$ implies $V = 0$, i.e. the particle 
becomes on-shell. One can show that the spectrum is that of gluons emitted
from on-shell quarks when $V = 0$. $k_0$ is the energy of the soft gluon emitted
at angle $\theta$ with the parent quark whose velocity is $\beta$ and energy is $E$.
The virtuality is replaced by $V=\sqrt{q^2-m^2}=\sqrt{E^2-p^2-m^2}$. 
The emitted gluon carry a fraction of parent parton energy consistent with 
$|\vec{p_i}|sin\theta>>k_0$ (for dominance of soft gluon emission). 
Different limits of $F$ given in Eq.~\ref{deadvirt} will be worth exploring.

\noi (i) For zero virtuality ($V=0$) of the massive quark, Eq.~\ref{deadvirt} 
reduces to the conventional dead cone factor~(Fig. \ref{figkmpaul}): 
\begin{figure}[h]      
\begin{center}                                                               
\includegraphics[scale=0.4]{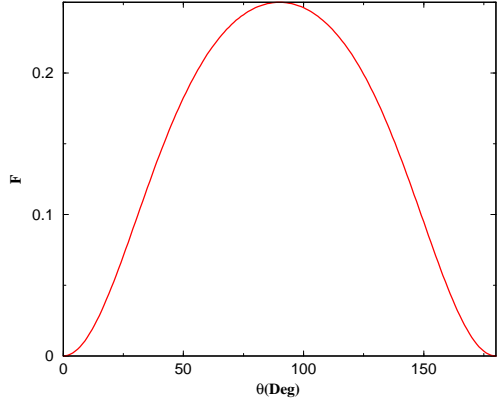}                                 
\end{center}                                                                 
\caption{Plot of the spectrum $F$ in $V=0$ limit (Eq.~\ref{kharvirt}), $m=1.5$ GeV, $\beta=0.5$}
\label{figkmpaul}                                                               
\end{figure}                                                                 

\bea
F
\longrightarrow
\frac{\beta^2 sin^2 \theta}
{(1-\beta^2 cos^2 \theta)^2}
\label{kharvirt}
\eea
This is the well-known conventional dead cone for a gluon emitted by a 
massive quark for large angles. The divergence of the factor is shielded by the quark mass
or virtuality through $\beta (<1)$.  In fact, for on-shell quarks with small $\theta$ 
one can show that Eq.~\ref{kharvirt} (see~\cite{khar}) boils down to:
\be
F=\frac{1}{\theta^2}\frac{1}{(1+\theta_0^2/\theta^2)^2},\,\,\,\, \theta_0=m/E
\ee
and 
for highly virtual quarks (large $V$) the $F$ can be written as:
\be
F=\frac{\omega^2}{E^2}
\ee

\noi(ii)Now,  the light quark limit ($\beta=1$) can be investigated. 
For $V=0,~\beta=1$, 
\be
F\sim\frac{1}{sin^2 \theta}
\label{light}
\ee
For light quarks Eq.~\ref{light} ensures the absence of dead-cone suppression at $\theta=0$ and $\pi$ 
for vanishing virtuality.

\begin{figure}[h]      
\begin{center}                                                               
\includegraphics[scale=0.6]{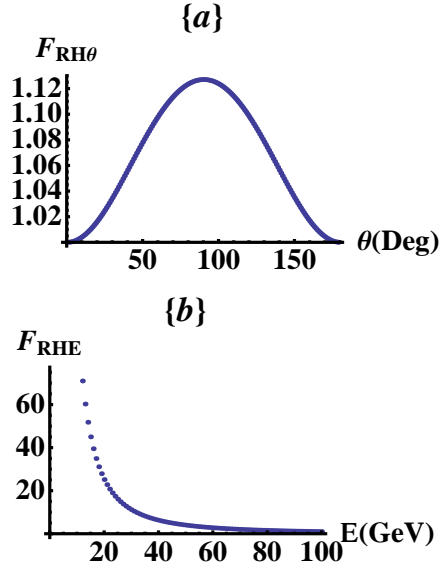}                                 
\end{center}                                                                 
\caption{(Colour online) The variation of (a)$F_{\mathrm RH\theta}(E,\theta)$,
defined in the text with $\theta$ for $E=1.5$ GeV,
(b) with $E$ for $\theta=\pi/4$ for virtual heavy quarks. 
The results displayed in this figure are evaluated for heavy quark mass, 
$m=1.27$ GeV, emitted gluon energy, $k_0=20$ MeV and $\beta=0.5$. 
The quark virtuality is defined as $V^2=E^2-\beta^2E^2-m^2$. 
}
\label{fig4}                                                               
\end{figure}                                                                 
\begin{figure}[h]      
\begin{center}                                                               
\includegraphics[scale=0.6]{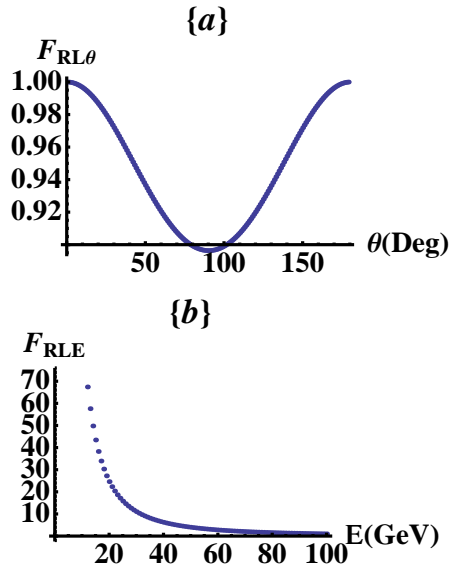}                                 
\end{center}                                                                 
\caption{(Colour online) The variation of (a)$F_{\mathrm RL\theta}(E,\theta)$,
defined in the text  with $\theta$ for $E=3$ GeV,
(b) with $E$ for $\theta=\pi/4$ for virtual light partons. 
The results displayed in this figure are evaluated for vanishing quark mass, 
$\beta=0.95$ and emitted gluon energy, $k_0=10$ MeV. 
The quark virtuality is defined as $V^2=E^2(1-\beta^2)$. 
}
\label{fig5}                                                               
\end{figure}                                                                 
In Fig.~\ref{fig4} the suppression of the energy loss, $F$ for heavy quarks is displayed. The 
variation of $F_{\mathrm RH\theta}=F(E=1.5 GeV,\theta)/F(E=1.5 GeV,\theta=0)$ 
with $\theta$ is depicted in Fig.~\ref{fig5}(a). It is interesting to note that
for vanishingly small virtuality the suppression is similar to that of a 
conventional dead cone that appear for massive on-shell quarks (Fig~\ref{figkmpaul}).  
In Fig.~\ref{fig4} (b) the variation of $F_{\mathrm RHE}=F(E,\theta=\pi/4)/F(E=100 GeV,\theta=\pi/4)$ with $E$ 
is displayed for heavy quarks. For large  virtuality (which increases with parton energy, $E$) 
the suppression is large. 

Fig.~\ref{fig5} illustrates the suppression of the energy loss for light quarks. 
In Fig.~\ref{fig5} (a) the variation of $F_{\mathrm RL\theta}=F(E=3 GeV,\theta)/F(E=3 GeV,\theta=0)$ with $\theta$ 
is shown for  light partons. It is important to note that the variation of 
$F_{\mathrm RL\theta}$ with $\theta$ for light quark with low virtuality is drastically different from 
the corresponding quantity, $F_{\mathrm RH\theta}$ for heavy quark. 
This is obvious because for low virtuality the light 
partons are not subjected to any dead cone suppression at $\theta=0$ and $\pi$ 
unlike heavy quarks. Moreover, the $sin^{-2}\theta$ behaviour for light quarks (Eq.~\ref{light})
ensures a minimum at $\theta=\pi/2$ as opposed to a maximum at the same $\theta$ for heavy quarks. 
In Fig.~\ref{fig5} (b) the variation of $F_{\mathrm RLE}=F(E,\theta=\pi/4)/F(E=100 GeV,\theta=\pi/4)$ with $E$ 
is depicted. 
We note that the suppression is large for high $E$ 
and the behaviour of $F_{\mathrm RLE}$
is  similar to $F_{\mathrm RHE}$
which indicates that the suppression due to virtuality overwhelm the 
effects  due to the conventional dead cone.  

\begin{figure}[h]      
\begin{center}                                                               
\includegraphics[scale=0.4]{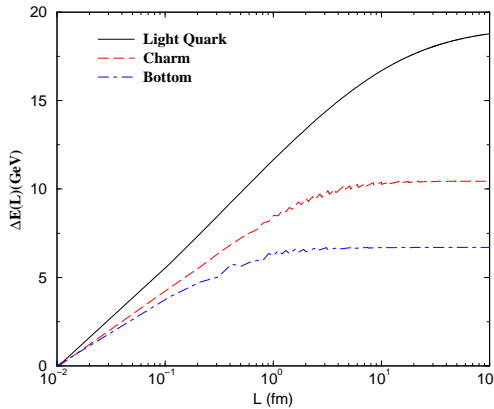}                                 
\end{center}                                                                 
\caption{Variation of energy loss, $\Delta E$ as a function of
path traversed, $L$ in vacuum. The energy of the quark is taken as $E=20$ GeV.}
\label{fig6}                                                               
\end{figure}

The energy loss ($\Delta E(L)$) of quarks  as a function of
path length ($L$) traversed by the off-shell parton in vacuum 
can be evaluated (see ~\cite{kop} for details)  
by using the emitted gluon spectrum given by Eq.~\ref{deadvirt}. 
The results are displayed in Fig.~\ref{fig6}. 
We note that the energy loss of light and heavy quarks differ significantly
at large path length or time when the propagating quarks acquire enough field to radiate.
However, at small path length the value of $\Delta E$ for light and heavy quarks 
are similar.

When we want to talk about radiative energy loss of high energy partons
we mean the absorption of radiation given off is absorbed in the medium. 
So, we must take into account the interaction of emitted radiation with
the medium. Consequently, the dispersion relation of the emitted gluon should change.
This change in gluon dispersion relation is encoded in the thermal quark self energy;
and as the inverse of imaginary part of self energy gives radiation
production rate, the formation time of radiation is also modified. This
effect of modified dispersion relation of emitted gluon (or photon)
is called Ter-Mikaelian (TM) effect~\cite{klein}. The QCD analogue 
of TM effect is discussed in ~\cite{kampfertermikaelian,dglv}. ~\cite{gossiaux}
implements TM effect by replacing $x^2m^2$ by $x^2m^2+(1-x)m_D^2$, where $m_D$
is thermal mass of gluon. However, ~\cite{kampfertermikaelian} repeats 
the single, double and multiple scattering calculations of ~\cite{wgp}
by assuming a modified dispersion relation, $k^2={\omega_0^2(T)}$, of emitted gluon.
$\omega_0$ parametrizes gluon self energy in medium. \cite{kampfertermikaelian}
shows that in the phase space region where abelian radiation does not occur,
the gluon radiation is suppressed due to polarization properties of the medium.
\cite{dglv,djotermikaelian} extends the results of \cite{kampfertermikaelian}
using 1-Loop HTL self-energy of gluons.  \cite{djotermikaelian} shows
a $\sim$30\% decrease in 1st order in opacity fractional energy loss for heavy quarks
of 10 GeV energy when TM effect is taken into account. Whereas, \cite{gossiaux}
shows $\sim$1.5 times increase in radiative energy spectra per unit length 
when gluon thermal mass is taken into consideration. Such changes indicate
that one can barely neglect the effect of gluon interaction with medium
while probing QGP.

So far, we have talked about the energy loss of HQs in an infinite medium.
But, how does the energy loss depend on the size of medium? \cite{baieretal} 
shows that length dependent energy loss $\Delta E(L) \propto L^2$, when we 
are considering the coherent region, {\it i.e.} the emitted gluon energy 
is soft and is within the factorization and the Bethe-Heitler 
limits (see \cite{annrev}). When gluon energy ($k_0$) is of the order of that of parent parton ($E$),
the energy loss per unit length becomes independent of length of medium. This is
the limit ($k_0\sim E$) when evaluation of transport coefficients without addressing
to size of medium works more efficiently. 

There is also a path integral approach of radiative energy loss proposed in \cite{zakharov}. 
Path integral approach is shown to be equivalent to the approach of \cite{baieretal} 
in \cite{annrev}. An alternate formalism (GLV) proposed in \cite{glv,glvnpb} performs 
a systematic expansion in opacity (the mean number of jet scatterings).
Opacity is quantitatively given by $\bar{n}=L/\lambda$, where $L$ is the target thickness and
$\lambda$ is the mean free path. One analytic limit applies to plasmas
where mean number of scattering is small~\cite{gw}. The other
limit applies to thick plasma where $\bar{n}>>1$~\cite{baieretal}. 

\vskip 0.2in

\section*{Acknowledgements:} SM and TB are supported by DAE, Govt. of India.
SKD acknowledges the support by the ERC StG under the QGPDyn Grant n. 259684.
The authors thank Jan-e Alam and Md. Younus for fruitful discussions
and critical reading of the manuscript. The authors
also take the opportunity to thank the anonymous reviewers for
their relevant suggestions which have undoubtedly helped to increase the 
quality of manuscript.

\end{document}